\documentclass[twocolumn]{svjour3}

\smartqed
\usepackage{graphicx}
\usepackage{mathptmx}
\usepackage[normalem]{ulem}
\usepackage{subfigure}
\usepackage[ugly]{units}
\usepackage{amssymb,amsmath,bm,bbm,booktabs}
\usepackage{flushend}

\newtheorem{LEO}{LEO}
\newtheorem{PEO}{PEO}
\newtheorem{Lemma}[LEO]{Lemma}
\newtheorem{Proposition}[PEO]{Proposition}

\journalname{Pattern Analysis and Applications}

\begin{document}

\title{Parametric and Nonparametric Tests for Speckled Imagery}
\author{Renato J.\  Cintra \and Alejandro C.\ Frery \and Abra\~ao D.\ C.\ Nascimento}
\institute{R.\ J.\  Cintra \at
            Department of Statistics\\
            Federal University of Pernambuco\\
            Recife, Pernambuco, Brazil\\
            \email{rjdsc@de.ufpe.br}\\
            \and
            A.\ C.\ Frery \at
            CPMAT \& LCCV\\
     		   Instituto de Computa\c c\~ao\\
            Universidade Federal de Alagoas\\
            Macei\'{o}, Alagoas, Brazil\\
            \email{acfrery@pesquisador.cnpq.br}
\and 
 	   A.\ D.\ C.\ Nascimento \at
            Statistics Postgraduate Program\\
            Federal University of Pernambuco\\
            Recife, Pernambuco, Brazil\\
            \email{abraao.susej@gmail.com}           
            }
\date{Received: date / Accepted: date}
\maketitle

\begin{abstract}
Synthetic aperture radar (SAR) has a pivotal role as a remote imaging method.
Obtained by means of co\-he\-rent illumination, SAR images are contaminated with speckle noise.
The statistical modeling of such contamination is well described according with the multiplicative model and its implied $\mathcal{G}^0$ distribution.
The understanding of SAR imagery and scene element identification is an important objective in the field.
In particular, reliable image contrast tools are sought.
Aiming the proposition of new tools for evaluating SAR image contrast, we investigated new methods based on stochastic divergence.
We propose several divergence measures specifically tailored for $\mathcal{G}^0$ distributed data.
We also introduce a nonparametric approach based on the Kol\-mo\-go\-rov-Smir\-nov distance for $\mathcal{G}^0$ data.
We devised and assessed tests based on such measures, and their performances were quantified according to their test sizes and powers.
Using Monte Carlo simulation, we present a robustness analysis of test statistics and of maximum likelihood estimators for several degrees of innovative contamination.
It was identified that the proposed tests based on triangular and arithmetic-geometric measures outperformed the Kol\-mo\-go\-rov-Smirnov methodology.
\end{abstract}

\section{Introduction}\label{sec:int}

Synthetic aperture radar (SAR), ultrasound-B, sonar, and la\-ser images are obtained with coherent illumination and, consequently, their appearance is affected by a signal-de\-pen\-dent granular noise called spe\-ckle~\cite{OliverandQuegan1998}.
Due to deviations from classical properties of additivity and Gaussian distribution, speckled images require tailored processing and analysis techniques which could conveniently rely on the statistical properties of the data.

The ability to discriminate SAR image regions is an important tool for identifying and classifying distinct targets in the scene.
Such capability has also been employed as a validation criterion for novel methodologies in SAR image change detection~\cite{IngladaMercier2007}.
These facts have motivated the proposal of contrast measures based on the statistical properties of image data~\cite{FreryCorreiaFreitas:ClassifMultifrequency:IEEE:2007,KarouiandFabletandBoucherandPieczynskiandAugustin2008,InformationContrastAnalysisPolarimetric}.
Therefore, an accurate modeling of speckled data is a major step for SAR image analysis~\cite{FreitasFreryCorreia:Environmetrics:03}.
Indeed, the probability distribution that describes the image data is a crucial information for any subsequent statistical analysis.

Among the proposals for SAR image modeling, the multiplicative model~\cite{FreitasFreryCorreia:Environmetrics:03} has proven to be an important framework for the statistical processing and analysis of SAR data~\cite{EstimationEquivalentNumberLooksSAR}.
Assuming that the hypotheses suggested by the multiplicative model are valid, the $\mathcal{G}^0$ distribution arises as an efficient and effective probabilistic model for speckled data~\cite{FreryandCribariNetoandSouza2004,StatisticalModelingSARImagesSurvey,mejailfreryjacobobustos2001,MejailJacoboFreryBustos:IJRS,VasconcellosandFreryandSilva2005}.

Hypothesis test methods have been considered as a venue to quantifying the image contrast between different regions in SAR imagery.
Edge detection~\cite{ParametricNonparametricEdgeDetectionSpeckledImages,GambiniandMejailandJacobo-BerllesandFrery}, classification~\cite{HoekmanandQuinones2000,Manolovaandanne2008}, target identification~\cite{HRRAutomaticTargetRecognition}, change detection~\cite{ConditionalCopulasChangeDetection} and oil spill identification~\cite{TextureEntropyOilSpillRADARSAT} rely on parametric and nonparametric hypothesis tests.
Although some nonparametric methods are computationally convenient, they seldom take explicitly the roughness information into account~\cite{ParametricNonparametricEdgeDetectionSpeckledImages}.
As a consequence, such simple approaches are unable to provide any information about that important parameter.
Additionally, these methods cannot offer any known statistical property that could be utilized for the design of hypothesis testing pro\-ce\-du\-res~\cite{GambiniandMejailandJacobo-BerllesandFrery}.
Thus, the proposal of hypothesis testing methods based on roughness dependent contrast measures is a much sought tool for speckled data analysis.

In recent years, the interest in adapting information-theo\-retic tools to image processing has increased notably.
In~\cite{InformationContrastAnalysisPolarimetric}, the Bhattacharyya and (symmetrized) Kullback-Leibler distances were applied as scalar contrast measures for polarimetric and interferometric SAR imagery.
The works~\cite{FreryNascimentoCintraICIP,HypothesisTestingSpeckledDataStochasticDistances} proposed several parametric methods based on $(h,\phi)$-di\-ver\-gen\-ce measures for speckled data.

Often regarded as a classical benchmark tool, the Kol\-mo\-go\-rov-Smir\-nov distance has been used for segmentation and labeling of SAR images~\cite{philipe}.
This nonparametric test has found applications in polarimetric speckled data analysis~\cite{HoekmanandQuinones2000}.
Additionally, the Kol\-mo\-go\-rov-Smir\-nov distance was also applied as a goodness-of-fit measure for SAR data modeling~\cite{DabaandBell1994}.

Whereas on the one hand a careful statistical modeling is fundamental, on the other hand nature is more complex than most models.
Even sophisticated statistical models and tools may have their effectiveness diminished when data deviate from the assumed underlying hypotheses, even mildly.
Robust statistics is a way of tackling this problem, and resistant techniques have been used with success in the SAR literature~\cite{AllendeFreryetal:JSCS:05,BustosFreryLucini:Mestimators:2001,FrerySantAnnaMascarenhasBustos:ASP:98,KerstenandLeeandAinsworth2005}.

This paper presents three contributions regarding the analysis of SAR imagery with stochastic measures of contrast.
Firstly, it quantifies the ability of selected parametric methods based on divergences measures by means of the size and the power of associated hypothesis tests using Monte Carlo.
Such quantities were compared to the ones related to the nonparametric Kolmogorov-Smirnov test. 
It also presents an expression for the Kol\-mo\-go\-rov-Smir\-nov distance between $\mathcal{G}^0$ distributions.
Secondly, it employs a contamination model for representing `innovative outliers'~\cite{Fox1972} and assesses the behavior of maximum likelihood (ML) estimators under the $\mathcal{G}^0$ model in the presence of contamination.
Finally, it submits hypothesis tests to different degrees of contamination, aiming to investigate the robustness of their sizes.
Additionally, it presents an application to actual SAR data.

The remainder of the paper is organized as follows.
Section~\ref{sec:mm} discusses the multiplicative model and the $\mathcal{G}^0$ distribution. 
The hypothesis tests under consideration, namely those derived from $(h,\phi)$-divergences and the Kolmogorov-Smirnov distance, are derived in Section~\ref{sec:distance}.
Section~\ref{Simulation} presents Monte Carlo experiments in order to assess ML estimation and hypothesis tests under situations which include ``pure'' and contaminated data.
Actual SAR data is analyzed in Section~\ref{sec:real}.
Section~\ref{sec:Conc} concludes the paper.

\section{The multiplicative model and the $\mathcal{G}^0$ distribution}\label{sec:mm}

Meaningful statistical processing of SAR imagery is always associated to a statistical model.

In~\cite{StatisticalModelingSARImagesSurvey}, Gao provided a comprehensive exposition of existings models.
In particular, three major categories of models were identified: (i) the multiplicative model~\cite{FreitasFreryCorreia:Environmetrics:03,freryetal1997a}, (ii)~models based on the generalized central limit theorem~\cite{KuruogluandZerubia2004}, and~(iii) the empirical distribution model~\cite{Andai2009777}. 
Other approaches include the joint distribution model~\cite{Blake1995}, the mixed Gaussian model~\cite{DoulgerisandEltoft2010}, mixtures of Gamma laws~\cite{FiniteGammaMixtureModelingMinimumMessageLengthInference}, and the correlation based model~\cite{leeetal1994b}.

The multiplicative model has a phenomenological nature which is closely tied to the physics of the image formation~\cite{OliverandQuegan1998}.
Zhang~\cite{Zhang2005} showed evidence that the distributions stemming from the product between independent random variables which describe the speckle noise and the backscatter outperform other models.
In \cite[p. 776]{StatisticalModelingSARImagesSurvey}, Gao includes the following note about the multiplicative model:
\begin{quote}
The most attractive achievement among them is the statistical modeling on extremely heterogeneous region of SAR images proposed by Frery~\textit{et al.}~\cite{freryetal1997a} who [\dots] has introduced the original idea that for the purpose of statistical modeling, SAR images can be divided into homogeneous regions, heterogeneous regions and extremely heterogeneous regions, according to their contents.
\end{quote}
Such assertion, related to $\mathcal G^0$ distribution, is due to the following facts:
\begin{enumerate}
\item it has as many parameters as the $K$ law, i.e., scale, roughness and number of looks, and they can also be interpreted, leading to scene understanding;
\item the roughness is expressive, in the sense that it is able to describe more targets than the $K$ law~\cite{mejailfreryjacobobustos2001};
\item its density does not depend on special functions, as is the case of the $K$ law~\cite{freryetal1997a};
\item its cumulative distribution function can be obtained using its relationship with the Snedekor's $F$ distribution~\cite{MejailJacoboFreryBustos:IJRS}.
\end{enumerate}
We, therefore, admit the $\mathcal G^0$ law as the model for the observed data, regardless the target and, without loss of generality, in its intensity format, i.e., quadratic detection.

Consider positive and independent random variables $X$ and $Y$ as models for the terrain backscatter and the speckle noise, respectively.
The model for the return is $Z = X \cdot Y$.

The generalized inverse Gaussian distribution is proposed as a general model for the intensity backscatter.
However, a tractable particular case is the reciprocal gamma law~\cite{FreitasFreryCorreia:Environmetrics:03}, whose density function is given by
\begin{equation} \label{modelmultiplicative1}
f_{X}(x;\alpha,\gamma)=\frac{\gamma^{-\alpha}}{\Gamma{(-\alpha)}} x^{\alpha-1} \exp{\Big(-\frac{\gamma}{x}\Big)},\quad -\alpha,\gamma,x > 0.
\end{equation}

In intensity format, the random variable $Y$ obeys the gamma distribution with unitary mean~\cite{FreitasFreryCorreia:Environmetrics:03}, which has the following density 
\begin{equation} \label{modelmultiplicative2}
f_{Y}(y;L)=\frac{L^L}{\Gamma(L)}y^{L-1}\exp(-Ly), \quad y > 0, L\geq1,
\end{equation}
where $L$ is the number of looks.
Throughout this paper, the number of looks is assumed to be known and constant over the whole image.
A detailed account of the until recently largely unexplored issue of estimating $L$ is provided in~\cite{EstimationEquivalentNumberLooksSAR}.

Considering independent random variables with the densities presented in Equations~\eqref{modelmultiplicative1} and~\eqref{modelmultiplicative2}, we obtain that the density of $Z$ is expressed by
\begin{align} \label{modelmultiplicative3}
f_{Z}(z;\alpha,\gamma,L)&= \frac{L^L \Gamma{(L-\alpha)}}{\gamma^\alpha\Gamma{(-\alpha)} \Gamma{(L)}} z^{L-1} (\gamma+Lz)^{\alpha-L}, \nonumber \\  
&=\frac{\Gamma(L-\alpha)}{\Gamma(-\alpha)\Gamma(L)}
\Big(\frac{L}{\gamma}\Big)^L \frac{z^{L-1}}{\bigl(1+\frac{L}{\gamma}z\bigr)^{L-\alpha}}, 
\end{align}
where $z>0$, $\alpha<0$ is the roughness, $\gamma>0$ is the scale, and $L\geq 1$.
We indicate this situation as $Z \sim \mathcal{G}^0(\alpha,\gamma,L)$.

As shown in~\cite{mejailfreryjacobobustos2001,MejailJacoboFreryBustos:IJRS}, this distribution can be used as an universal model for speckled data.
The $r$th moment of $Z$ is expressed by
\begin{equation} \label{modelmultiplicative4}
\operatorname{E}(Z^r)=\Big(\frac{\gamma}{L}\Big)^r \frac{\Gamma{(-\alpha-r)}}{\Gamma{(-\alpha)}} \frac{\Gamma{(L+r)}}{\Gamma{(L)}},
\end{equation}
if $-r>\alpha$. 
Otherwise, it is infinite.

Reparametrizing Equation~\eqref{modelmultiplicative3} with $M=-\alpha>0$ and $\gamma=M\mu$, we obtain the density of the Fisher model~\cite{Tisonetal2004} whose density is given by
\begin{align*}
f_{Z}(z; M,\mu,L)=\frac{\Gamma(L+M)}{\Gamma(M)\Gamma(L)}\Big(\frac{L}{M\mu}\Big)^L\frac{z^{L-1}}{\bigl(1+\frac{L}{M\mu}z\bigr)^{L+M}}.
\end{align*}
Using some Mellin transform properties for Fisher distribution, Galland~\textit{et al.}~\cite{Gallandetal2009} derived the so-called log-cummulants $w_i=\operatorname{E}(Z^i \log Z)$:
\begin{align*}
w_0=&\log \frac{M\mu}{L}+\psi^{0}(L)-\psi^{0}(M),\\
w_1=&\frac{M\mu}{M-1}\Big[ \log\frac{M\mu}{L}+\psi^{0}(L+1)-\psi^{0}(M-1)\Big], \\
w_2=&\frac{M^2(L+1)}{(M-1)(M-2)L}\Bigl[ \log\frac{M\mu}{L} +\psi^{0}(L+2)-\psi^{0}(M-2) \Bigr],
\end{align*}
where $\psi^{0}(\cdot)$ is the digamma function.
As the Fisher distribution is a reparametrization of the $\mathcal{G}^0$ distribution, the results in~\cite{Gallandetal2009,Tisonetal2004} apply to the latter.

\begin{figure}[ht]
\centering 
\includegraphics[width=.9\linewidth]{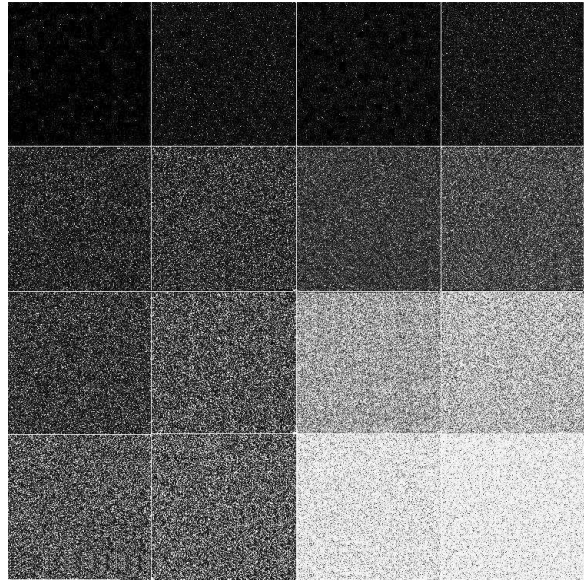} 
\caption{Synthetic $\mathcal{G}^0$ imagery for $(\mu,L) \in \{(1,1),(1,8),(10,1),(10,8)\}$ (left to right) and $\alpha \in \{-15,-8,-5,-3\}$ (top to bottom).}
\label{figure1}
\end{figure}

\begin{figure}[ht]
\centering
\subfigure[Varying $\alpha$ and $L$ for $\mu=1$\label{figure21}]{\includegraphics[width=.48\linewidth]{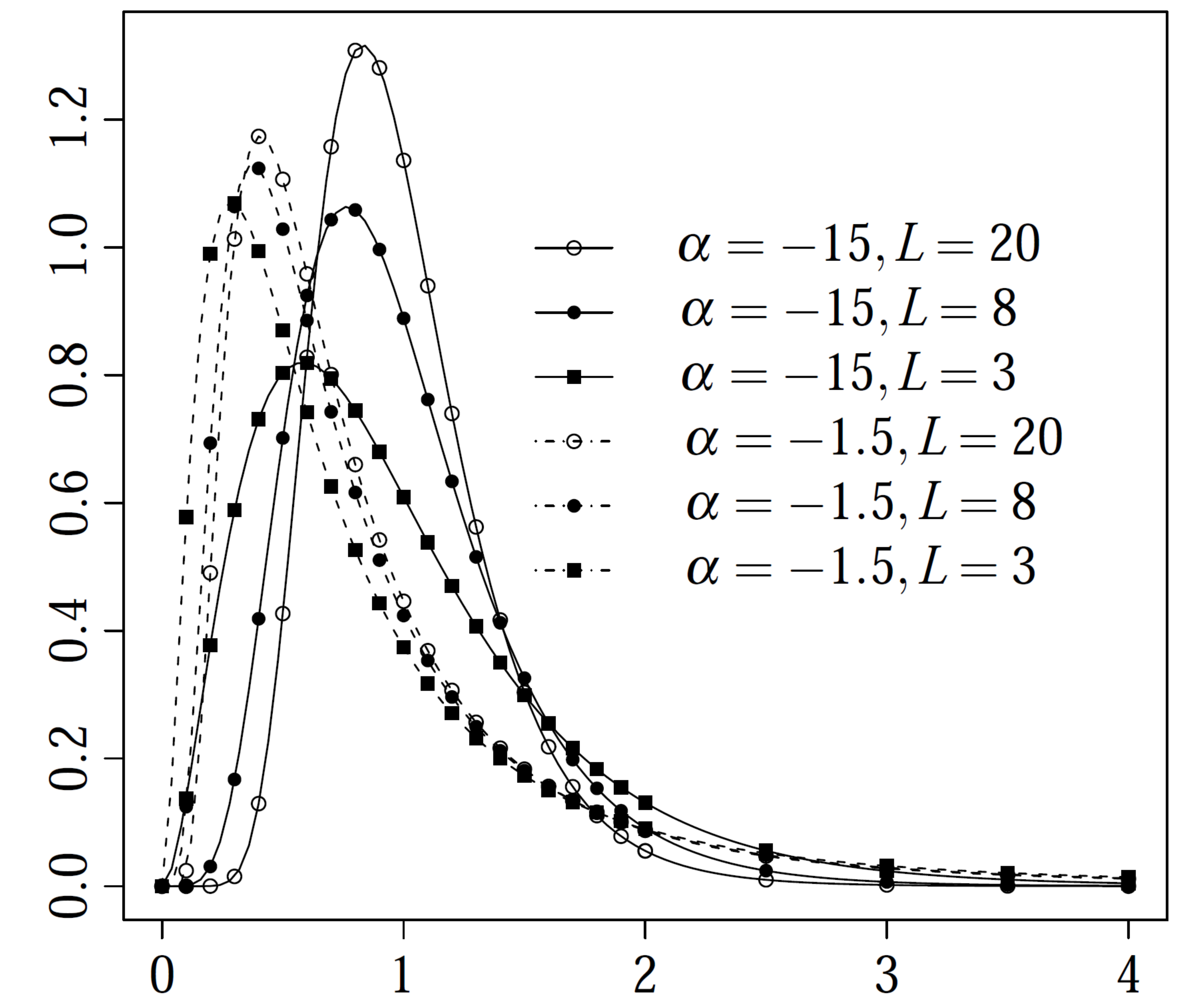}}%
\subfigure[Varying $\mu$ for $\alpha=-2.5$ and $L=3$ \label{figure22}]{\includegraphics[width=.48\linewidth]{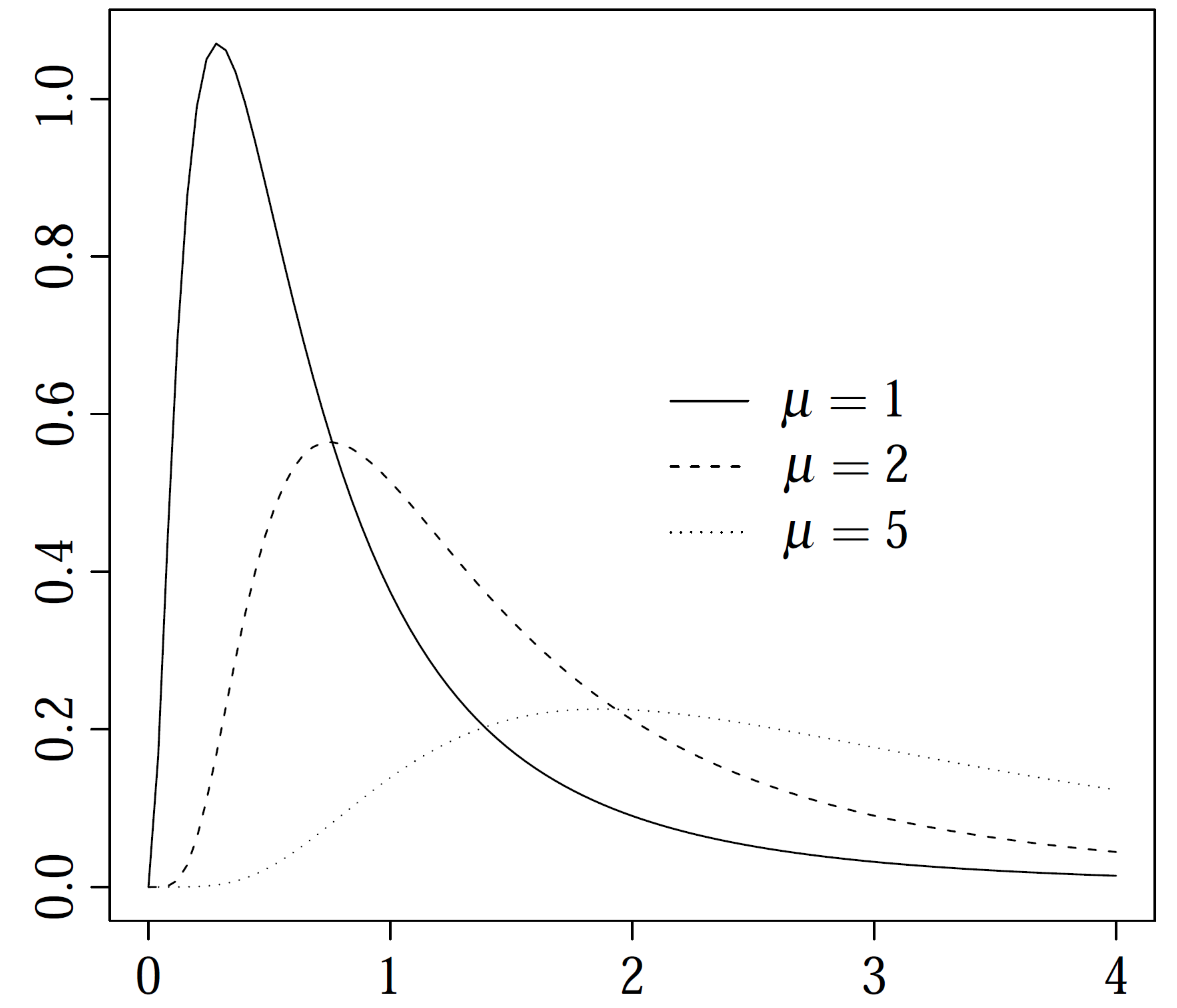}}%
\caption{$\mathcal{G}^0$ densities.}
\label{figure2}
\end{figure}

Fig.~\ref{figure1} shows sixteen $256 \times 256$ pixel regions of simulated speckled data according to the $\mathcal{G}^0$ distribution for $L=\{1,8\}$, $\alpha \in \{-15,-8,-5,-3\}$, and $\operatorname{E}(Z)=\mu \in \{1,10\}$.
The homogeneity depends on $\alpha$, the roughness parameter, while $\mu$ controls the mean brightness.
Fig.~\ref{figure21} and~\ref{figure22} show the density of $\mathcal{G}^0$ distributions with varying $(\alpha,L) \in \{(-1.5,20),(-1.5,8),$
$(-1.5,3),(-15,20),(-15,8),(-15,3)\}$, and $\mu \in \{1,2,5\}$ for $\mu=1$ and $(\alpha,L)=(-2.5,3)$, respectively.
Increasing $L$ or $\mu$ flattens the densities, while small values of $\alpha$ lead to images of  low variability.
 
Parameters $\alpha$ and $\gamma$ can be estimated according to several methods, including bias-reduced procedures~\cite{CribariFrerySilva:CSDA,SilvaCribariFrery:ImprovedLikelihood:Environmetrics,VasconcellosandFreryandSilva2005} and ro\-bust te\-chni\-ques (such as L- and M-estimators adapted to asymmetric distributions)~\cite{AllendeFreryetal:JSCS:05,BustosFreryLucini:Mestimators:2001}. 
Small sample methods are particularly relevant in image processing and analysis, for instance, in the determination of the equivalent number of looks, in adaptive filters and for training classifiers.
In such cases, samples are usually very small when compared to the full image~\cite{FreryNascimentoCintraICIP}, ranging from a single observation to, typically, less than a hundred.

ML estimation was adopted due to its optimal asymptotic properties~\cite{salicruetal1994}.
Let $Z_1,\ldots,Z_n$  be $n$ iid $\mathcal{G}^0(\alpha,\gamma,L)$ distributed random variables.
The ML estimator for $(\alpha,\gamma)$, for a given $L$, namely $(\widehat{\alpha},\widehat{\gamma})$, is any solution of the following system of
non-linear equations~\cite{HypothesisTestingSpeckledDataStochasticDistances}:
\begin{equation*}
\begin{array}{r}
\psi^{0}(-{\widehat\alpha})-\psi^{0}(L-{\widehat\alpha})-\log\widehat\gamma+\frac1n\sum_{i=1}^n \log( \widehat\gamma+L Z_i)=0,\\
-\frac{{\widehat\alpha}}{{\widehat\gamma}}+\frac{{\widehat\alpha}-L}{n}\sum_{i=1}^n({\widehat\gamma} + L Z_i)^{-1}=0.
\end{array}
\label{equationsystem}
\end{equation*}
In general, the above system of equations does not possess a closed form solution and numerical optimization methods are required.
We used the Broy\-den-Flet\-cher-Gold\-farb-Sha\-nno (BFGS) me\-thod, which is regarded as an accurate technique~\cite{Cribari--Netozarkos1999}.
Next section presents several contrast measures for $\mathcal G^0$ distributed images based on the above estimators.

\begin{figure}[htb]
\centering
\subfigure[SAR image with selected regions\label{MLestimatenew1}]{\includegraphics[width=.90\linewidth]{imagewithregions3paaN}}\\
\subfigure[Urban region\label{MLestimatenew2}]{\includegraphics[width=.49\linewidth]{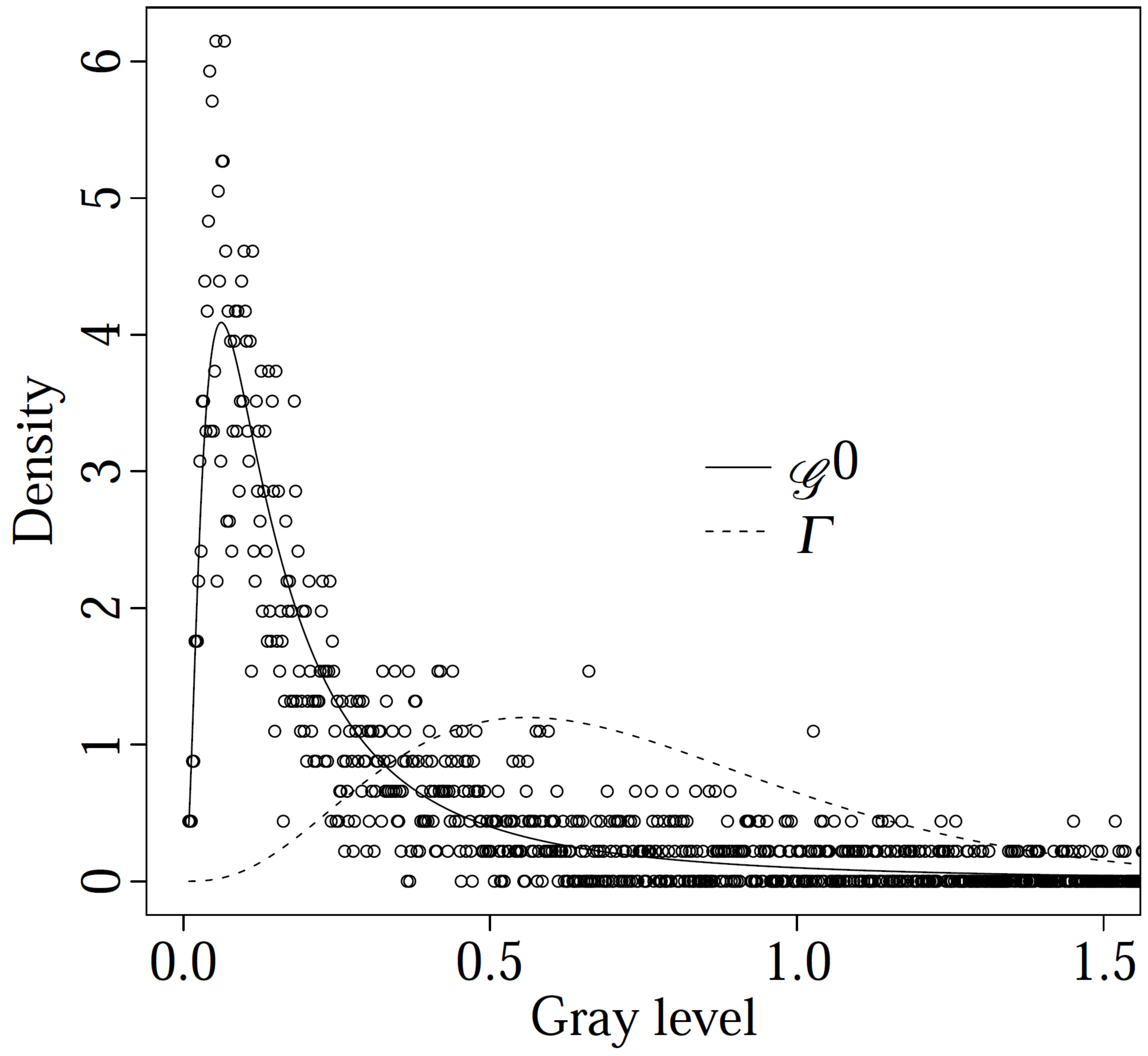}}
\subfigure[Forest region\label{MLestimatenew3}]{\includegraphics[width=.49\linewidth]{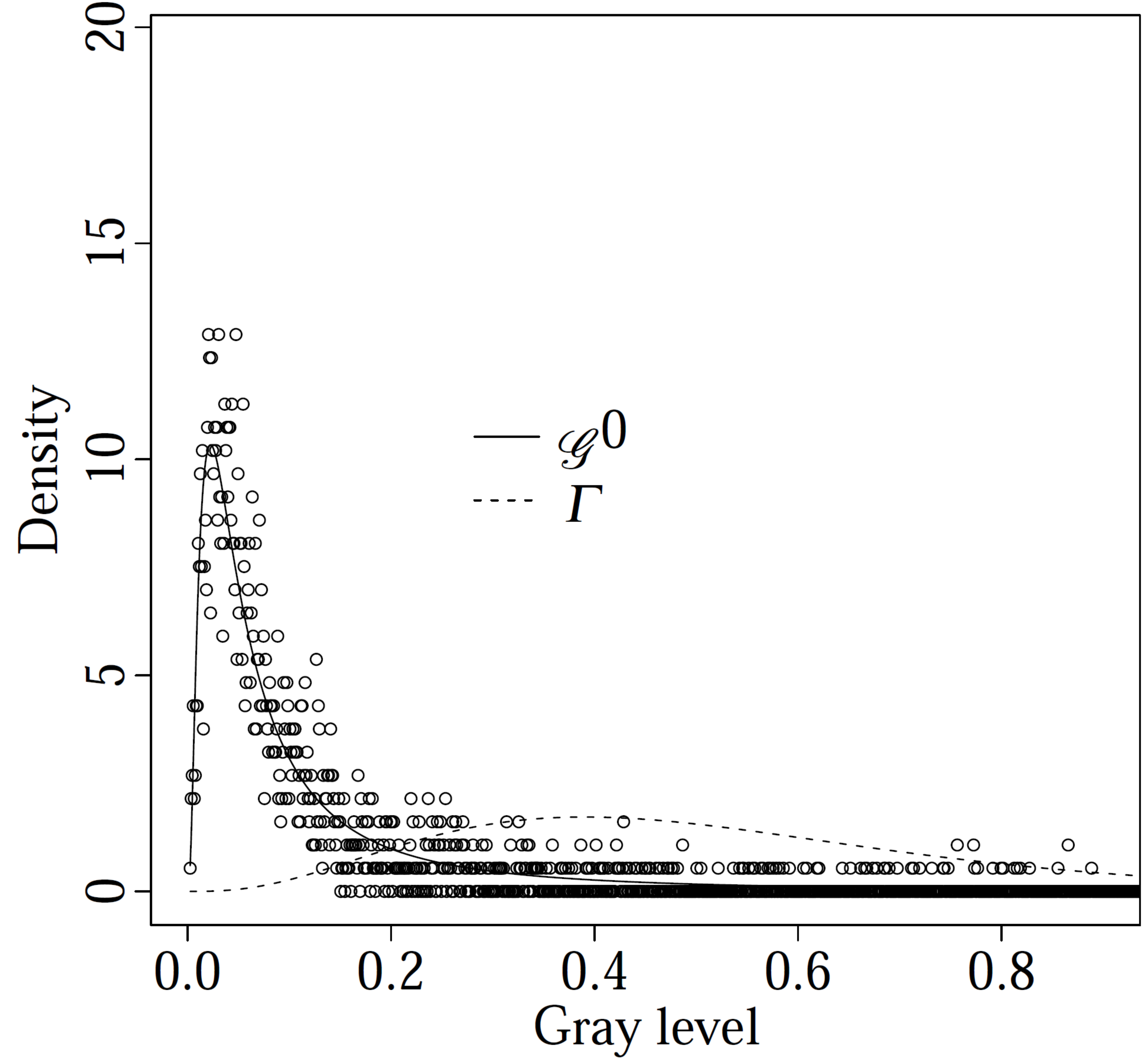}}
\caption{Real SAR image with selected regions  and empirical and fitted densities for differents urban and forest regions.}
\label{MLestimatenew}
\end{figure}

Fig.~\ref{MLestimatenew1} shows an image over San Francisco, CA, obtained by the AIRSAR sensor at band~L with four (nominal) looks. 
Urban and forest regions were selected, and Table~\ref{tabelapplica0} presents the sample sizes and ML estimates of $\alpha$ and $\gamma$.
Moreover, this table shows the sum of squared errors ($\operatorname{SSE}$) between the histogram ${f}_n$ and the fitted density $\widehat{f}_n$, which is given by~\cite{Zhangetal2009}
$$
\operatorname{SSE}=\sum_{n=1}^{\text{\# pixels}}\frac{(\widehat{f}_n-f_n)^2}{\text{\# pixels}}.
$$
Two densities are considered: the basic gamma model (the scaled version of the density presented in Equation~\eqref{modelmultiplicative2}) and the $\mathcal G^0$ distribution.

\begin{table}[htb]
\centering
\scriptsize 
\caption{ML estimates and SSE}\label{tabelapplica0}
\begin{tabular}{c r r@{.}l r   rrc}
\toprule
\multicolumn{5}{c}{} & \multicolumn{2}{c}{$\operatorname{SSE}$} & \\ \cmidrule(lr{.25em}){6-7} 
Regions & \multicolumn{1}{c}{$\overline{Z}$} &\multicolumn{2}{c}{$\widehat{\alpha}$} & \multicolumn{1}{c}{$\widehat{\gamma}\text{ }\;(\times 10^{-2})$} & $\mathcal G^0$ & $\Gamma$ & \# pixels \\ \midrule
Urban & 0.374 &$-$1&370 &  19.550 & 0.062 & 0.318 & 2277  \\
Forest & 0.260 & $-$1&275 & 7.461 & 0.108 & 0.389 & 1813  \\
\bottomrule
\end{tabular}
\end{table}

The values of SSE provide evidence that the $\mathcal G^0$ distribution outperforms the gamma distribution for the considered regions. 
The strongest difference occurs in urban regions, i.e., in situation with intense roughness, as presented in Figs.~\ref{MLestimatenew2}-\ref{MLestimatenew3}. 
As discussed in~\cite{mejailfreryjacobobustos2001}, among other references, the $\mathcal G^0$ law includes the gamma distribution as a particular case and, therefore, it is able to describe homogeneous targets as pasture, crops, and sea.

\section{Contrast measures for the $\mathcal{G}^0$ distribution}\label{sec:distance}

An image can be interpreted as a set of regions des\-cri\-bed by possibly different probability laws, and contrast analysis addresses the pro\-blem of quan\-ti\-fying how dis\-tin\-guishable two image regions are.
Many statistical ap\-proa\-ches have been developed to reach this goal, leading to the difficult problem of specifying accurate, expressive, and tractable metrics.
In general terms, these methods can be either pa\-ra\-me\-tric or nonparametric.

In a previous work, we proposed a class of pa\-ra\-me\-tric measures based on stochastic distances.
Advancing that study, among the existing nonparametric methods, we now include an approach based on the Kolmogorov-Smirnov test.
This method is reported to possess good discriminatory properties~\cite{LimandJuJang:2002} and has found applications in several instances~\cite{HoekmanandQuinones2000,philipe}.
In the following, we briefly describe the parametric methods.
Moreover, we adapt the Kolmogorov-Smir\-nov test to $\mathcal{G}^0$ distributed data.

We assume that $Z'$ and $Z''$ are $\mathcal G^0$ dis\-tri\-bu\-ted random variables with cumulative distribution functions $F_{Z'}$ and $F_{Z''}$, respectively.
These distributions share the support $\mathbb R_+$ and are defined over the same probability space, equipped with densities $f_{Z{'}}(z;\bm{\theta}_1)$ and $f_{Z''}(z;\bm{\theta}_2)$, respectively, where $\bm{\theta}_1$ and $\bm{\theta}_2$ are parameter vectors.

Section~\ref{CONTR} presents contrast measures derived from stochastic distances, while Section~\ref{KKS} discusses the use of the Kol\-mo\-go\-rov-Smir\-nov distance for the $\mathcal G^0$ model.

\subsection{Contrast measures based on divergence}\label{CONTR}

Information theoretical tools known as diver\-gence mea\-su\-res of\-fer en\-tro\-py ba\-sed me\-tho\-ds for con\-tras\-ting sto\-chas\-tic distributions.
Di\-ver\-gen\-ce measures were submitted to a systematic and compre\-hensive treatment and, as a result, the class of $(h,\phi)$-diver\-gences was proposed~\cite{salicruetal1994}.

The $(h,\phi)$-divergence between $f_{Z'}$ and $f_{Z''}$ with common support $I$ is defined by
\begin{eqnarray*} \label{generaldivergence}
D_{\phi}^h(Z',Z'')=h\Bigl(\int_{I} \phi\Bigl( \frac{f_{Z'}(z;\bm{\theta}_1)}{f_{Z''}(z;\bm{\theta}_2)}\Bigr) f_{Z''}(z;\bm{\theta}_2)\mathrm{d}z\Bigr),
\end{eqnarray*}
where $\phi\colon (0,\infty)\rightarrow[0,\infty)$ is a convex function, $h\colon(0,\infty)\rightarrow[0,\infty)$ is a strictly increasing function with $h(0)=0$, and indeterminate forms are assigned value zero.
Table~\ref{tab-1} shows the $h$ and $\phi$ func\-tions considered in this work.

\begin{table*}[hbt]
\centering   
\caption{($h,\phi$)-distances, statistics, $h$ and~$\phi$ functions, and related quantity}\label{tab-1}
\begin{tabular}{rcccc}
\toprule
{ $(h,\phi)$-{distance}} & Statistic & { $h(y)$} & { $\phi(x)$} & $v=[h{'}(0) \phi{''}(1)]^{-1}$\\
\cmidrule(lr{.25em}){1-1} \cmidrule(lr{.25em}){2-2} \cmidrule(lr{.25em}){3-3} \cmidrule(lr{.25em}){4-4} \cmidrule(lr{.25em}){5-5}
Kullback-Leibler ($d_\text{KL}$) &$S_{\text{KL}}$ & ${y}/{2}$ & $(x-1)\log x$  & $1$\\
Triangular ($d_\text{T}$)  &$S_{\text{T}}$ &  $y,\;0\leq y <2$ & $\frac{(x-1)^2}{x+1}$ & $1$ \\
Bhattacharyya ($d_\text{B}$)  &$S_{\text{B}}$ & $-\log(-y+1),0\leq y<1$ & $-\sqrt{x}+\frac{x+1}{2}$ & $4$\\
Arithmetic-geometric ($d_\text{AG}$) &$S_{\text{AG}}$ & $y$ & $\frac{x+1}{2}\log\frac{x+1}{2x}+\big(\frac{x-1}{2}\big)$  & $4$\\
\bottomrule
\end{tabular}
\end{table*}

These divergences are often asymmetric, and in order to solve this issue and to turn them into distances, a new measure $d_{\phi}^h$ can expressed as
\begin{equation}
d_{\phi}^h(Z',Z'')=\frac{D_{\phi}^h(Z',Z'') + D_{\phi}^h(Z'',Z')}{2}.
\label{eq:hphidistance}
\end{equation}

We will be interested in comparing samples from possibly the same distribution $\mathcal D(\bm{\theta})$, so instead of indexing Equation~\eqref{eq:hphidistance} by the random variables, we will do it by the parameters $\bm{\theta}_1$ and $\bm{\theta}_2$.
Furthermore, denote  $\widehat{\bm{\theta}_1}$ and $\widehat{\bm{\theta}_2}$ the maximum likelihood estimators of $\bm{\theta}_1$ and $\bm{\theta}_2$ based on independent samples of sizes $m$ and $n$, respectively.

\begin{Lemma}\label{prop-chi}
Let the regularity conditions pro\-po\-sed in \cite[p.~380]{salicruetal1994} hold.
If $\frac{m}{m+n} \xrightarrow[m,n\rightarrow\infty]{} \lambda\in(0,1)$ and $\bm{\theta}_1=\bm{\theta}_2$, then
\begin{eqnarray*}
\frac{2 mn}{m+n}\frac{1}{ h{'}(0) \phi{''}(1)}d^h_{\phi}(\widehat{\bm{\theta}_1},\widehat{\bm{\theta}_2})   \xrightarrow[m,n\rightarrow\infty]{\mathcal{D}}\chi_{M}^2,
\end{eqnarray*}
where ``$\xrightarrow[]{\mathcal{D}}$'' denotes convergence in distribution.
\end{Lemma}

Based on Lemma~\ref{prop-chi}, the following statistical hypothesis test for the null hypothesis $\bm{\theta}_1=\bm{\theta}_2$ can be derived:
\begin{align*}
S_{\phi}^h(\widehat{\bm{\theta}_1},\widehat{\bm{\theta}_2})=\frac{2mnv}{m+n}d^h_{\phi}(\widehat{\bm{\theta}_1},\widehat{\bm{\theta}_2}).
\end{align*}
Table~\ref{tab-1} shows the notation we for each test statistic, along with the values of $v=[h{'}(0) \phi{''}(1)]^{-1}$. 

We are now in position to state the following result.
\begin{Proposition}\label{p-3}
Let $m$ and $n$ be large, and consider the outcome $S_{\phi}^h(\widehat{\bm{\theta}_1},\widehat{\bm{\theta}_2})=s$.
The null hypothesis $\bm{\theta}_1=\bm{\theta}_2$ can be re\-jec\-ted at a nominal le\-vel $\eta$ if $\Pr\bigl(\chi^2_{M} > s  \bigr)\leq \eta$. 
\end{Proposition}

\subsection{Kolmogorov-Smirnov contrast measure}\label{KKS}

The empirical function of the sample $\bm Z=(Z_1,\dots,Z_n)$ of a random variable $Z$ is given by $\widehat{F}_{\bm Z}(z) = n^{-1} \#\{j: Z_j\leq z\}$.
Empirical functions can be used to test the null hypothesis $H_0\colon F_{Z'}=F_{Z''}$ against the alternative $H_1\colon F_{Z'}\neq F_{Z''}$ with random samples $\bm Z'$ and $\bm Z''$ of sizes $m$ and $n$, respectively, using the test statistic~\cite{Smirnov1933}
\begin{eqnarray*}
S_\text{KS}(\bm Z', \bm Z'')=\sqrt{\frac{mn}{m+n}} \underbrace{\max_{\substack{-\infty < z <\infty}}|\widehat{F}_{\bm Z'}(z)-\widehat{F}_{\bm Z''}(z)|}_{\widehat{d}_\text{KS}},
\end{eqnarray*}
where $\widehat{d}_\text{KS}$ is the empirical estimate for the Kolmogorov-Smirnov distance $d_\text{KS}$.
In this case, the analytic expression for $d_\text{KS}$ under $\mathcal G^0$ distributions with same number of looks is given by
\begin{align*}
d_\text{KS}(\bm Z', \bm Z'')&= \displaystyle \max_{\substack{-\infty < z <\infty}}\Big|L^L z^L \Big[  \nonumber\\
&\frac{\Gamma(L-\alpha_1)}{\gamma_1^L \Gamma(-\alpha_1)}{}_2F_1\Big(L,L-\alpha_1,L+1,\frac{-Lz}{\gamma_1} \Big)- \nonumber\\    
&\frac{\Gamma(L-\alpha_2)}{\gamma_2^L \Gamma(-\alpha_2)}{}_2F_1\Big(L,L-\alpha_2,L+1,\frac{-Lz}{\gamma_2} \Big) \Big]\Big|,
\end{align*}
where ${}_2F_1$ is the hypergeometric function.

\begin{Proposition}\label{p-5}
Assume $S_\text{KS}(\bm Z', \bm Z'')=s$. The null hypothesis $F_{Z'}=F_{Z''}$ can be rejected at level $\eta$ if $\Pr[S_\text{{\rm KS}}(\bm Z', \bm Z'') > \delta] = 1 - L(s) \leq \eta$, where $L(s)=1-2\sum_{k=1}^{\infty}(-1)^{k-1}\exp{(-2 k^2 s^2)}$ and $\delta$ is a critical value given in \cite{Smirnov1933}, provided that $m$ and $n$ are large.
\end{Proposition}

In terms of image analysis, Propositions~\ref{p-3} and~\ref{p-5} offer meth\-ods to statistically refute the hypothesis that two samples obtained in different regions can be described by the same distribution.
Such assessment is relevant, for instance, when choosing input training samples in supervised learning procedures, or when proposing similarity-based techniques such as segmentation and classification.

Since the aforementioned results are asymptotic and valid only when the underlying hypothesis holds, the following section shows a Monte Carlo approach for their assessment with finite samples and under contamination.
Such assessment is needed since finite samples and outliers are unavoidable in practical situations.

\section{Validation}\label{Simulation}

This section presents a qualitative assessment of the test proposed for checking whether two samples are distributed according to the same law when finite size samples are used.
The proposed tests are submitted to uncontaminated data in Section~\ref{sec:simul}.
Subsequently, a contamination model based on innovative outliers, inspired by corner reflectors, is considered in Section~\ref{sec:simul1}.
The effect of contamination on the ML estimates of $\mathcal G^0$ parameters is also assessed.

\subsection{Analysis with Simulated Data}\label{sec:simul}

Speckled data simulated with the $\mathcal{G}^0$ law were employed to assess the discussed parametric and nonparametric test statistics.
The situations herein assessed were three levels of roughness ($\alpha\in \{-1.5,-3,-5\}$), three mean return va\-lu\-es ($\mu \in \{1,2,5\}$), and two va\-lu\-es of the number of looks ($L\in \{1,8\}$).
The mean $\mu=-\gamma/(1+\alpha)$ is directly related to the image brightness.
We performed the tests at the $1\%$ significance level using square windows of $7\times 7$, $9\times 9$, and $11\times 11$ pi\-xels. 
Therefore, the considered sample sizes were $N=m=n \in \{49,81,121\}$.

Empirical size and power of the proposed tests were sou\-ght as a means to measure and compare their per\-for\-man\-ces.
Mon\-te Car\-lo experiments under five different scenarios were used.
Considering two image regions specified by parameter vectors $(\alpha_1, \mu_1, L)$ and $(\alpha_2, \mu_2, L)$, the fol\-lo\-wing scenarios were set: (i)~$\alpha_1 = \alpha_2$, $\mu_1 = \mu_2$, (ii) $\alpha_1=\alpha_2$, $\mu_1\neq\mu_2$, (iii)~$\alpha_1 > \alpha_2$, $\mu_1 > \mu_2$, (iv)~$\alpha_1 > \alpha_2$, $\mu_1 < \mu_2$, and (v)~$\alpha_1\neq\alpha_2$, $\mu_1=\mu_2$.
Whereas the first situation corresponds to image regions of same roughness and brightness, scenario~(ii) accounts for images of same roughness but with different brightness. 
Images of both distinct roughness and brightness are captured by scenarios~(iii) and~(iv).
Additionally, the fifth scenario considers regions of different roughness and equal brightness.

\begin{figure*}[htb]
\centering
\includegraphics[width=.7\linewidth]{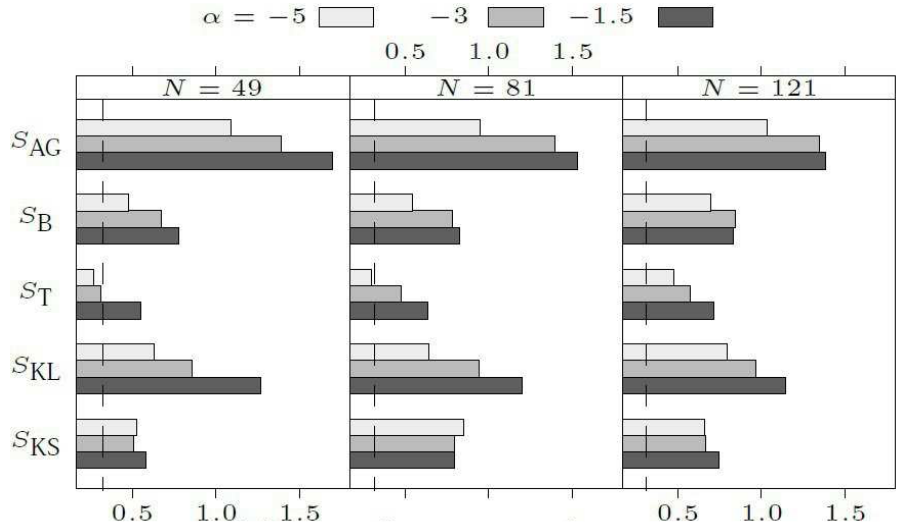}
\caption{Mean of empirical test size scaled by 10 at $1\%$ nominal level.}\label{size1}
\end{figure*}

Fig.~\ref{size1} shows the mean empirical test sizes, i.e., under scenario~(i).
With the exception of $S_\text{T}$, significance levels are above the nominal value (shown as a dashed line).
The tests tend to reject more often when $\alpha$ increases, with little effect from sample sizes.
In general, $S_\text{T}$ is the most efficient measure regarding the test size, since its estimative is in mean the closest to nominal values; $S_\text{T}$ is immediately followed by $S_\text{KS}$.
As will be seen later, $S_\text{KS}$ surprisingly loses this ability in the presence of contamination.

Goudail and R\'efr\'egier~\cite{Goudail2004} present graphically a re\-la\-tion\-ship between test power and the Kull\-ba\-ck and Bhat\-ta\-cha\-ryya dis\-tances.
Similarly, we illustrate the estimates of test powers with respect to the discrepancy measure given by
\begin{align*} 
\bar{D}= \bigr(d_\text{KL}+d_\text{T}+d_\text{B}+d_\text{AG}+d_\text{KS} \bigl)/5,
\end{align*}
where, for the sake of simplicity, arguments $(\widehat{\bm{\theta}_1}, \widehat{\bm{\theta}_2})$ were suppressed.

Tables~\ref{table-7}-\ref{table-32} present the test empirical power in scenarios~(ii)-(v), respectively. 
As a general behavior, the empirical power increases with the number of looks ($L$) and/or the sample size ($N$). 

\begin{figure*}[htb]
\centering
{\includegraphics[width=.65\linewidth]{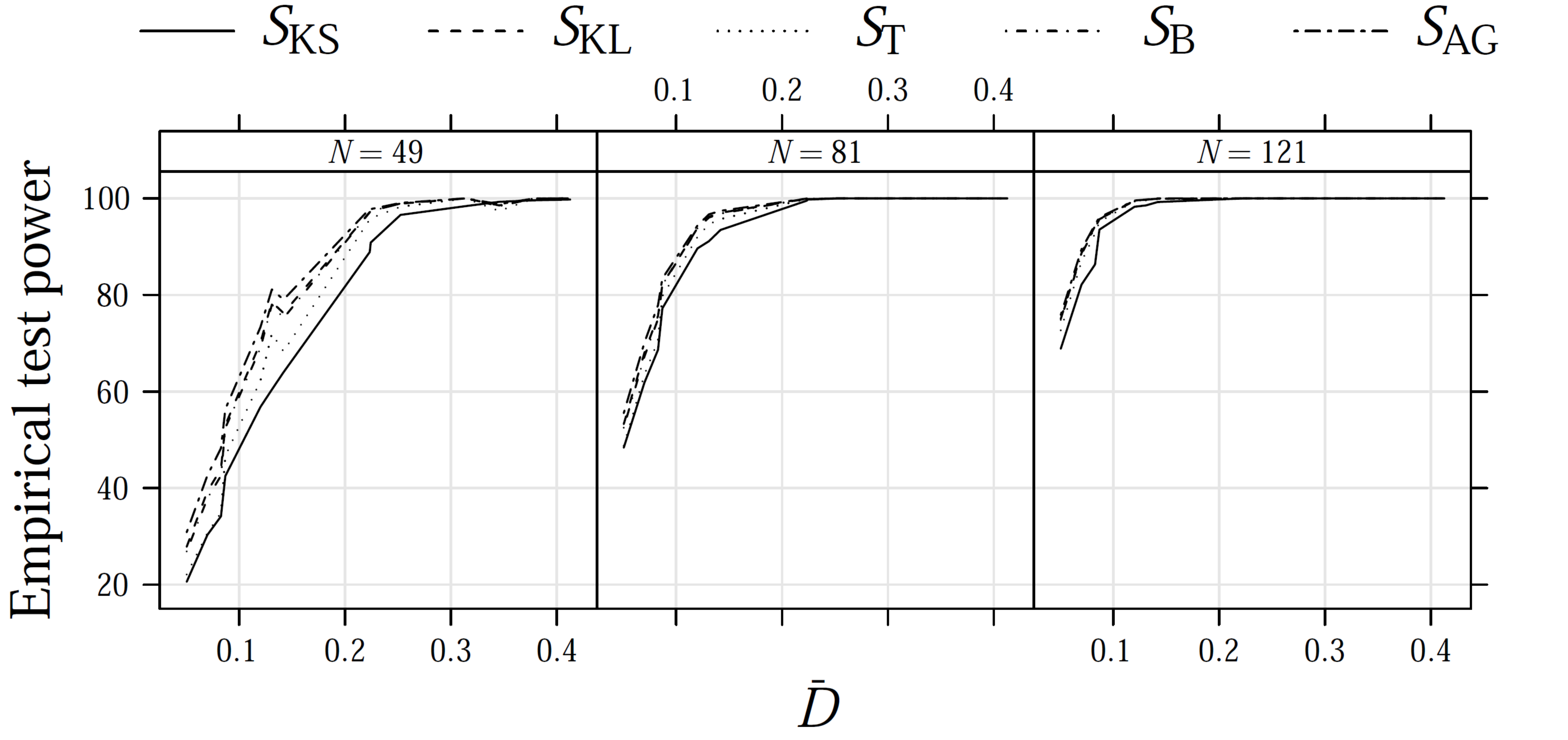}}
\caption{Empirical test power in situations (ii) for different contrast measure at $1\%$ nominal level.}
\label{power2}
\end{figure*}

Considering a fixed roughness, Table~\ref{table-7} provides evidence that the brightness change has more influence on the power in homogeneous regions.
Fig.~\ref{power2} illustrates the results of this table with respect to $\bar{D}$ for different sample sizes.
In this case, we noticed that parametric tests outperformed the Kolmogorov-Smirnov test.

\begin{figure*}[htb]
\centering
\subfigure[$\alpha_1=-1.5\text{ and }\alpha_2=-3.0$\label{poder1}]{\includegraphics[width=.39\linewidth]{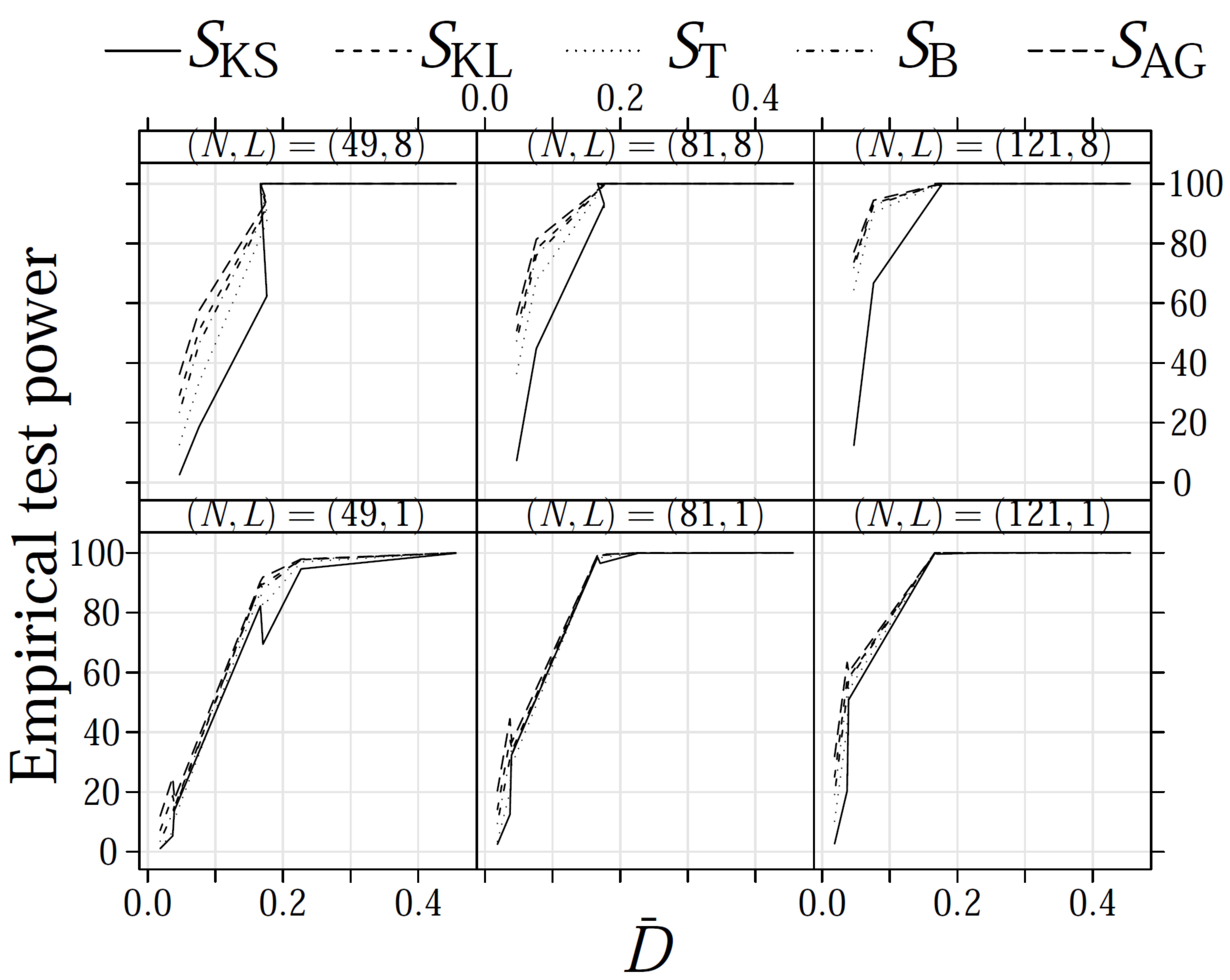}}
\subfigure[$\alpha_1=-1.5\text{ and }\alpha_2=-5.0$\label{poder2}]{\includegraphics[width=.39\linewidth]{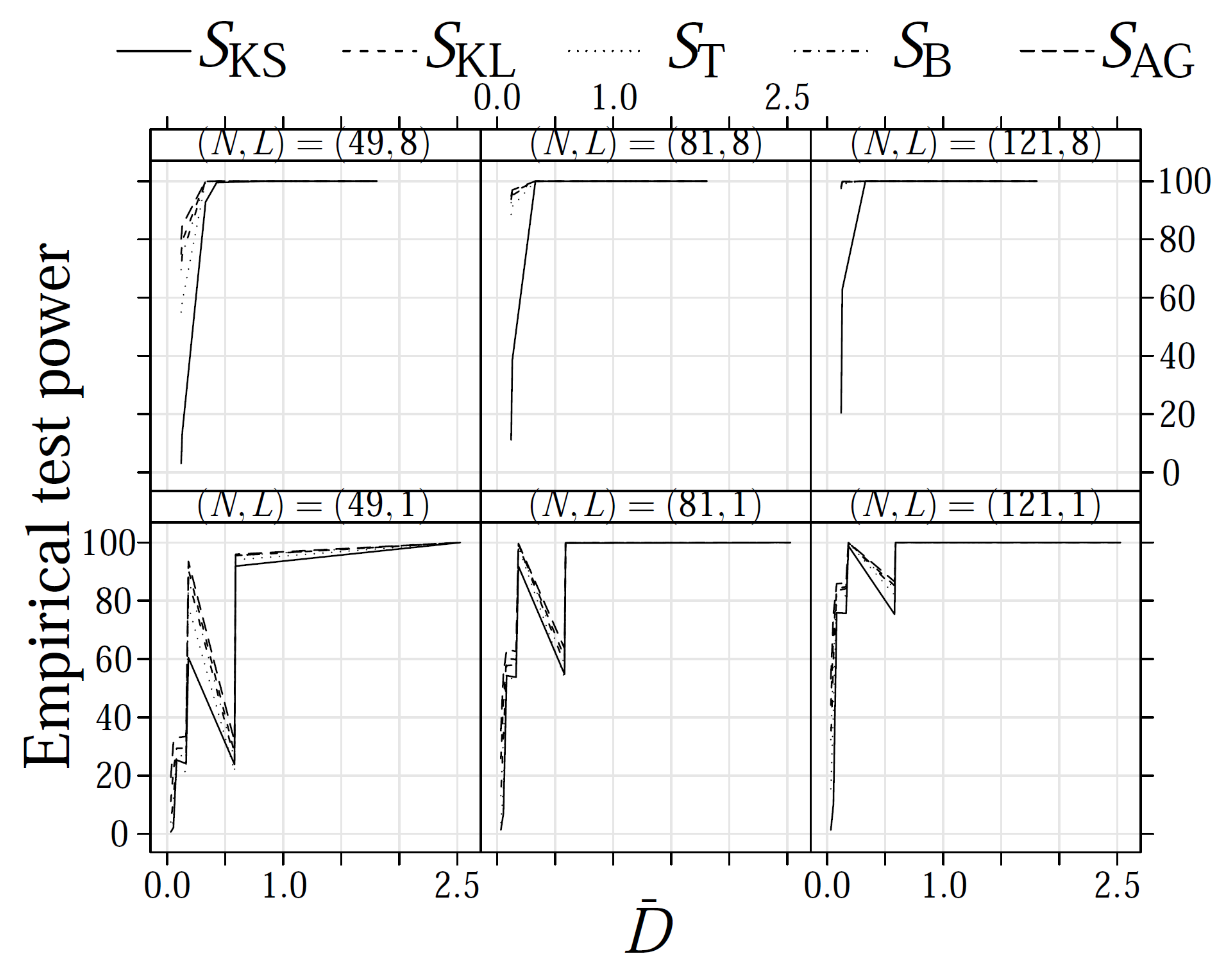}}\\
\subfigure[$\alpha_1=-3.0\text{ and }\alpha_2=-5.0$\label{poder3}]{\includegraphics[width=.39\linewidth]{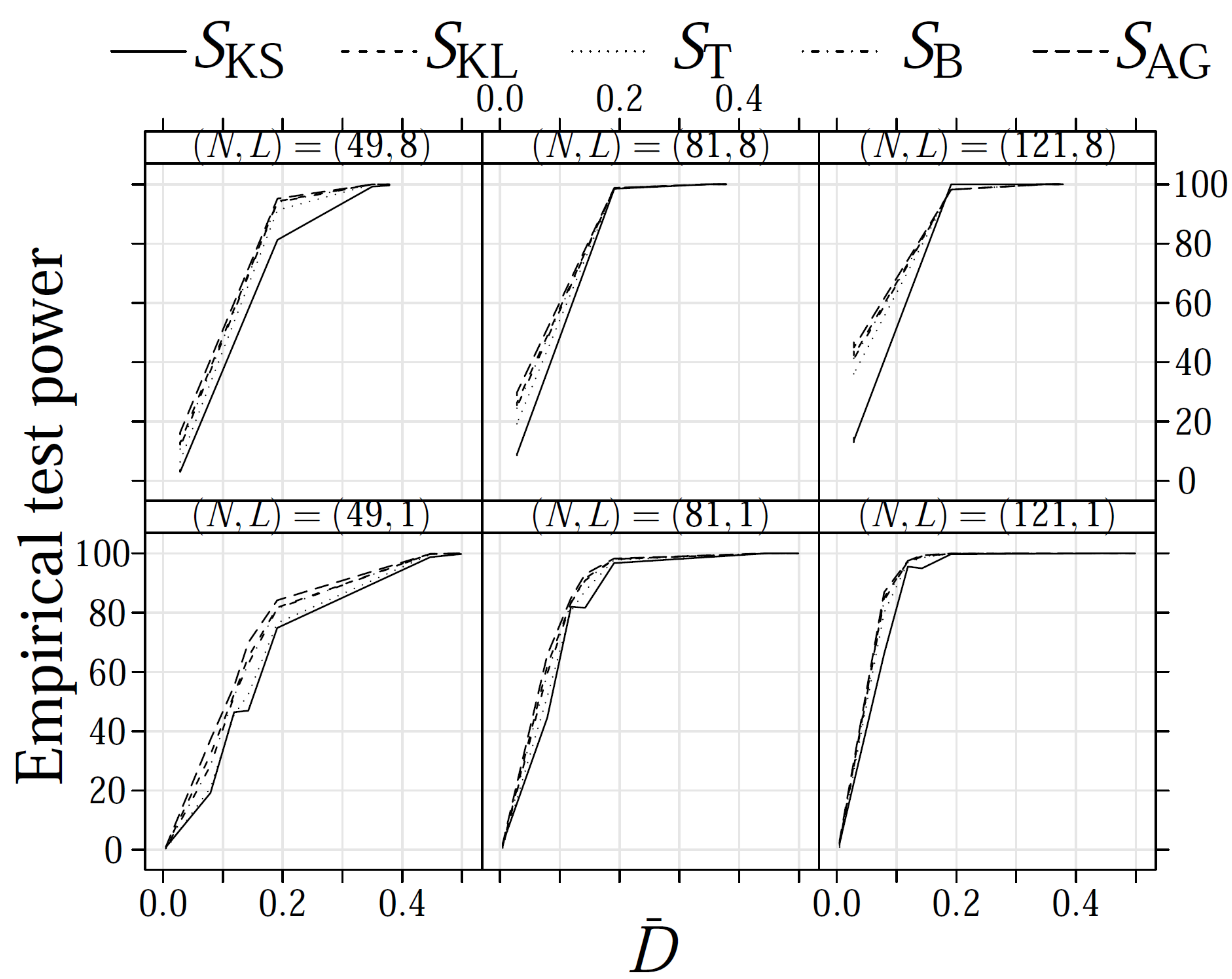}}
\caption{Empirical test power in situations (iii)-(v) for different contrast measure at $1\%$ nominal level.}
\label{poderes}
\end{figure*}

In Tables~\ref{table-3}, \ref{table-31}, and~\ref{table-32}, the inequality $\text{Power}(S_\text{KS})\leq \text{Power}(S_\text{T})\leq\text{Power}(S_\text{B})\leq\text{Power}(S_\text{KL})\leq\text{Power}(S_\text{AG})$ is satisfied in e\-ve\-ry case of scenarios~(iii) and~(iv). 
For scenario~(v), this fact holds when the number of looks is 8.
Fig.~\ref{poderes} presents the powers as functions of $\bar{D}$ for different roughness pairs. 
The power increases with the sample size and with $\bar{D}$, as expected.
These results show that the $S_\text{AG}$ test outperforms other techniques with respect to the test power, since it is able to discriminate more often samples from different distributions.

\begin{table*}[htb]                                                                                                                                                   
\centering \scriptsize                                                                                                                                                    \caption{Rejection rates of $(h,\phi)$-divergence and Kolmogorov-Smirnov tests for scenarios (ii)}                                                                        \label{table-7}                                                                                                                                                       
\begin{tabular}{c cc c r@{ } r@{ } r@{ } r@{ } r c r@{ } r@{ } r@{ } r@{ } r c r@{ } r@{ } r@{ } r@{ } r}\toprule 
$\mu_2/\mu_2$ & $L$ & N & $\alpha$ & $S_\text{KS}$ & $S_\text{KL}$ &  $S_\text{T}$ &  $S_\text{B}$&   $S_\text{AG}$ & $ \alpha $ & $S_\text{KS}$ &                    
$S_\text{KL}$ &  $S_\text{T}$ &  $S_\text{B}$&   $S_\text{AG}$ & $ \alpha $ & $S_\text{KS}$ & $S_\text{KL}$ &  $S_\text{T}$ &  $S_\text{B}$& $S_\text{AG}$\\ 
\cmidrule(lr{.25em}){1-1} \cmidrule(lr{.25em}){2-3} \cmidrule(lr{.25em}){4-4} \cmidrule(lr{.25em}){5-9} \cmidrule(lr{.25em}){10-10} \cmidrule(lr{.25em}){11-15} \cmidrule(lr{.25em}){16-16} \cmidrule(lr{.25em}){17-21}  
$2.0$  & $1$ & $49 $ &$-1.5$& 20.58	& 27.84	& 22.11	& 26.89	& 30.85	&$-3.0$& 30.31	& 39.34	& 30.70	& 37.87	& 42.63	&$-5.0$ & 34.16	& 43.97	& 34.89	& 42.58	& 48.26	\\
       &     & $81 $ &      & 48.36	& 53.24	& 48.79	& 52.51	& 55.52	&      & 61.85	& 68.03	& 63.52	& 67.26	& 70.14	&     & 68.64	& 75.65	& 70.50	& 75.01	& 77.89	\\               &     & $121$ &      & 68.85	& 75.04	& 72.68	& 74.84	& 75.85	&      & 82.13	& 88.64	& 86.89	& 88.42	& 89.42	&     & 86.33	& 94.30	& 92.73	& 93.93	& 94.58	\\ \cmidrule(lr{.25em}){2-3} \cmidrule(lr{.25em}){5-9} \cmidrule(lr{.25em}){11-15} \cmidrule(lr{.25em}){17-21}
       & $8$ & $49 $ &      & 60.40	& 78.53	& 72.00	& 78.02	& 81.15	&      & 90.85	& 97.41	& 95.89	& 97.34	& 97.78	&     & 98.40	& 100.00   & 100.00	  & 100.00   & 100.00	  \\ 
       &     & $81 $ &      & 91.13	& 96.20	& 94.65	& 96.04	& 96.69	&      & 99.75	& 99.84	& 99.82	& 99.84	& 99.84	&     & 100.00	  & 100.00   & 100.00	  & 100.00   & 100.00	  \\
       &     & $121$ &      & 98.55	& 99.73	& 99.62	& 99.73	& 99.75	&      & 100.00	  & 100.00	  & 100.00	  & 100.00	  & 100.00	  &     & 100.00	  & 100.00   & 100.00	  & 100.00   & 100.00	  \\ 
\cmidrule(lr{.25em}){1-1} \cmidrule(lr{.25em}){2-3}  \cmidrule(lr{.25em}){5-9}  \cmidrule(lr{.25em}){11-15}  \cmidrule(lr{.25em}){17-21}  
       
 $2.5$ & $1$ & $49 $ &      & 42.51	& 53.07	& 46.66	& 52.46	& 56.36	&      & 56.80	& 70.15	& 62.23	& 69.00	& 73.33	&     & 64.00	& 76.25	& 68.30	& 75.20	& 79.03	\\
       &     & $81 $ &      & 77.22	& 82.18	& 79.73	& 82.05	& 83.41	&      & 89.64	& 93.87	& 92.03	& 93.78	& 94.38	&     & 93.47	& 97.02	& 95.72	& 96.86	& 97.42	\\
       &     & $121$ &      & 93.53	& 95.73	& 95.13	& 95.68	& 96.08	&      & 98.27	& 99.54	& 99.40	& 99.50	& 99.56 &     & 99.24	& 99.93	& 99.88	& 99.93	& 100.00	  \\ \cmidrule(lr{.25em}){2-3} \cmidrule(lr{.25em}){5-9} \cmidrule(lr{.25em}){11-15} \cmidrule(lr{.25em}){17-21}
       & $8$ & $49 $ &      & 88.89	& 97.16	& 95.65	& 97.13	& 97.85	&      & 99.60	& 99.96	& 99.95	& 99.96	& 99.98 &     & 99.96	& 100.00	  & 100.00	  & 100.00	  & 100.00  	\\ 
       &     & $81 $ &      & 99.53	& 99.95	& 99.91	& 99.95	& 99.96	&      & 100.00	  & 100.00	  & 100.00  	& 100.00	  & 100.00	  &     & 100.00	  & 100.00	  & 100.00	  & 100.00	  & 100.00	  \\
       &     & $121$ &      & 100.00	  & 100.00	  & 100.00  	& 100.00	  & 100.00	  &      & 100.00	  & 100.00	  & 100.00  	& 100.00	  & 100.00	  &     & 100.00	  & 100.00	  & 100.00	  & 100.00	  & 100.00	  \\ 

\cmidrule(lr{.25em}){1-1} \cmidrule(lr{.25em}){2-3}  \cmidrule(lr{.25em}){5-9}  \cmidrule(lr{.25em}){11-15}  \cmidrule(lr{.25em}){17-21}  

$5.0$ & $1$ & $49 $ &      & 96.58	& 98.91	& 98.27	& 98.95	& 99.06	&      & 99.29	& 98.55	& 97.39	& 98.42	& 98.81	&     & 99.76	& 100.00	  & 100.00	  & 100.00	  & 100.00	  \\
       &     & $81 $ &      & 100.00	  & 100.00  	& 100.00  	& 100.00  	& 100.00  	&      & 100.00	  & 100.00	  & 100.00	  & 100.00	  & 100.00	  &     & 100.00	  & 100.00	  & 100.00	  & 100.00	  & 100.00	  \\
       &     & $121$ &      & 100.00	  & 100.00  	& 100.00  	& 100.00  	& 100.00  	&      & 100.00	  & 100.00	  & 100.00	  & 100.00	  & 100.00	  &     & 100.00	  & 100.00	  & 100.00	  & 100.00	  & 100.00	  \\ 
\cmidrule(lr{.25em}){2-3} \cmidrule(lr{.25em}){5-9} \cmidrule(lr{.25em}){11-15} \cmidrule(lr{.25em}){17-21}
       & $8$ & $49 $ &      & 100.00	  & 100.00	  & 100.00	  & 100.00	  & 100.00	  &      & 100.00	  & 100.00	  & 100.00	  & 100.00	  & 100.00	  &     & 100.00	  & 100.00	  & 100.00	  & 100.00	  & 100.00	  \\
       &     & $81 $ &      & 100.00	  & 100.00	  & 100.00	  & 100.00	  & 100.00	  &      & 100.00	  & 100.00	  & 100.00	  & 100.00	  & 100.00	  &     & 100.00	  & 100.00	  & 100.00	  & 100.00	  & 100.00	  \\
       &     & $121$ &      & 100.00	  & 100.00	  & 100.00	  & 100.00	  & 100.00	  &      & 100.00	  & 100.00	  & 100.00	  & 100.00	  & 100.00	  &     & 100.00	  & 100.00	  & 100.00	  & 100.00	  & 100.00	  \\
\hline                                                                                                                                                                
\end{tabular}                                                                                                                                                         
\end{table*}                                                                                                                                                          

\begin{table*}[htb]
\centering
\scriptsize
\caption{Rejection rates of $(h,\phi)$-divergence and Kolmogorov-Smirnov tests for scenarios (iii)-(v) for roughness pair $(\alpha_1,\alpha_2)=(-1.5,-3)$} \label{table-3}
\begin{tabular}{ r@{ }r@{ }c@{ }r@{ } r@{ } r@{ } r@{ } r@{ }c@{ }r@{ } r@{ } r@{ } r@{ } r@{ }c@{ }r@{ } r@{ } r@{ } r@{ } r@{ }}\toprule \multicolumn{2}{c}{Constants}&\multicolumn{1}{c}{$\gamma$}& \multicolumn{5}{c}{$\mu_1>\mu_2$} &\multicolumn{1}{c}{$\gamma$}& \multicolumn{5}{c}{$\mu_1=\mu_2$}&\multicolumn{1}{c}{$\gamma$}&\multicolumn{5}{c}{$\mu_1<\mu_2$} \\

\cmidrule(lr{.25em}){1-2} \cmidrule(lr{.25em}){3-3} \cmidrule(lr{.25em}){4-8} \cmidrule(lr{.25em}){9-9} \cmidrule(lr{.25em}){10-14} \cmidrule(lr{.25em}){15-15} \cmidrule(lr{.25em}){16-20}

$L$ & $N$ &$(\gamma_1,\gamma_2)$& $S_\text{KS}$ & $S_\text{KL}$ &  $S_\text{T}$ &  $S_\text{B}$&   $S_\text{AG}$ & $(\gamma_1,\gamma_2)$ & $S_\text{KS}$ & $S_\text{KL}$ &  $S_\text{T}$ &  $S_\text{B}$&   $S_\text{AG}$ & $(\gamma_1,\gamma_2)$ & $S_\text{KS}$ & $S_\text{KL}$ &  $S_\text{T}$ &  $S_\text{B}$& $S_\text{AG}$\\                 

\cmidrule(lr{.25em}){1-2} \cmidrule(lr{.25em}){3-3} \cmidrule(lr{.25em}){4-8} \cmidrule(lr{.25em}){9-9} \cmidrule(lr{.25em}){10-14} \cmidrule(lr{.25em}){15-15} \cmidrule(lr{.25em}){16-20}                                      
                                                       
1  & 49   &  (1,2)      &  1.05	   & 7.12	   & 1.17	   & 3.53	  & 11.94    & (0.5,2)  &   13.45   & 16.15   & 11.64   & 15.14   & 18.56   &    (0.5,4)    & 82.27  & 89.51    & 86.89    &  89.48    & 90.47   \\            
                & 81   &             &  2.47	   & 14.10   & 3.45	   & 9.46	  & 20.33    &          &   31.80   & 35.39   & 30.38   & 34.11   & 38.25   &               & 98.53  & 99.28    & 99.20    &  99.26    & 99.35   \\  
                & 121  &             &  2.64	   & 24.98   & 10.20   & 19.17	& 31.83    &          &   51.13   & 58.36   & 54.75   & 57.68   & 60.30	  &               & 99.89  & 100.00   & 100.00   &  100.00	 & 100.00  \\  

\cmidrule(lr{.25em}){1-2}  \cmidrule(lr{.25em}){4-8}  \cmidrule(lr{.25em}){10-14}  \cmidrule(lr{.25em}){16-20}                                                                                                                   
8  & 49   &             &  2.58	   & 29.18   & 12.68   & 23.52	& 36.20    &          &   62.35   & 92.41   & 87.65   & 91.71   & 94.06   &               & 99.98  & 100.00   & 100.00   &  100.00	 & 100.00  \\  
& 81   &             &  7.33	   & 50.77   & 36.49   & 47.39	& 56.12    &          &   93.31   & 99.51   & 99.16   & 99.45   & 99.53   &               & 100.00 & 100.00   & 100.00   &  100.00	 & 100.00  \\  
& 121  &             &  12.40   & 73.73   & 64.55   & 71.93	& 77.13    &          &   99.22   & 99.85   & 99.84   & 99.84   & 99.85   &               & 100.00 & 100.00   & 100.00   &  100.00	 & 100.00  \\    

\cmidrule(lr{.25em}){1-2} \cmidrule(lr{.25em}){3-3} \cmidrule(lr{.25em}){4-8} \cmidrule(lr{.25em}){9-9} \cmidrule(lr{.25em}){10-14} \cmidrule(lr{.25em}){15-15} \cmidrule(lr{.25em}){16-20}                                      

1  & 49   &  (2.5,2)    &  69.45   & 89.53   & 82.24   & 87.95	& 91.89    &  (1,4)   &   13.78   & 15.55   & 10.85   & 14.41   & 18.28   &    (0.5,10)   & 99.95  & 100.00   & 100.00   &  100.00	 & 100.00  \\    
& 81   &             &  96.51   & 99.24   & 98.39   & 99.17	& 99.45    &          &   31.65   & 35.47   & 30.65   & 34.15   & 37.74   &               & 100.00 & 100.00   & 100.00   &  100.00	 & 100.00  \\    
& 121  &             &  99.62   & 100.00  & 99.98   & 100.00	& 100.00   &          &   49.98   & 58.51   & 54.67   & 57.57   & 60.48   &               & 100.00 & 100.00   & 100.00   &  100.00	 & 100.00  \\  

\cmidrule(lr{.25em}){1-2}  \cmidrule(lr{.25em}){4-8}  \cmidrule(lr{.25em}){10-14}  \cmidrule(lr{.25em}){16-20}                                                            

8  & 49   &             &  99.49   & 100.00  & 100.00  & 100.00	& 100.00   &          &   62.35   & 92.19   & 87.23   & 91.47   & 94.03   &               & 100.00 & 100.00   & 100.00   &  100.00	 & 100.00  \\  
                & 81   &             &  100.00  & 100.00  & 100.00  & 100.00	& 100.00   &          &   92.93   & 99.55   & 99.29   & 99.49   & 99.62   &               & 100.00 & 100.00   & 100.00   &  100.00	 & 100.00  \\  
                & 121  &             &  100.00  & 100.00  & 100.00  & 100.00	& 100.00   &          &   99.47   & 99.98   & 99.96   & 99.98   & 100.00  &               & 100.00 & 100.00   & 100.00   &  100.00	 & 100.00  \\  

\cmidrule(lr{.25em}){1-2} \cmidrule(lr{.25em}){3-3} \cmidrule(lr{.25em}){4-8} \cmidrule(lr{.25em}){9-9} \cmidrule(lr{.25em}){10-14} \cmidrule(lr{.25em}){15-15} \cmidrule(lr{.25em}){16-20}                                      

1  & 49   &  (2.5,4)    &  5.29	   & 18.77   & 6.70    & 13.06	& 24.88    & (2.5,10) &   13.71   & 14.83   & 10.65   & 14.00   & 17.47   &    (1,10)     & 94.65  & 97.74    & 96.96    &  97.74    & 97.92   \\  
& 81   &             &  12.44   & 37.45   & 19.88   & 31.21	& 44.52    &          &   32.27   & 33.52   & 28.79   & 32.34   & 36.47   &               & 99.93  & 100.00   & 99.98    &  100.00	 & 100.00  \\    
& 121  &             &  20.35   & 57.55   & 41.69   & 53.04	& 63.38    &          &   50.87   & 58.23   & 54.08   & 57.37   & 60.09   &               & 99.98  & 100.00   & 100.00   &  100.00	 & 100.00  \\    

\cmidrule(lr{.25em}){1-2}  \cmidrule(lr{.25em}){4-8}  \cmidrule(lr{.25em}){10-14}  \cmidrule(lr{.25em}){16-20}                                                            

8  & 49   &             &  18.71   & 51.03   & 33.57   & 46.56	& 57.55    &          &   62.78   & 92.53   & 87.72   & 91.97   & 94.00   &               & 100.00 & 100.00   & 100.00   &  100.00	 & 100.00  \\    
& 81   &             &  44.84   & 78.12   & 67.69   & 76.02	& 81.37    &          &   92.11   & 99.58   & 99.25   & 99.56   & 99.69   &               & 100.00 & 100.00   & 100.00   &  100.00	 & 100.00  \\    
& 121  &             &  66.75   & 93.38   & 90.38   & 92.84	& 94.44    &          &   99.27   & 100.00  & 100.00  & 100.00  & 100.00  &               & 100.00 & 100.00   & 100.00   &  100.00	 & 100.00  \\    

\bottomrule                                                                                                                                                               \end{tabular}                                                                                                                                                             \end{table*}                                                                                                                                                                                                                     

\begin{table*}[htb]                                                                                                                                                       \centering                                                                                                                                                                \scriptsize                                                                                                                                                               \caption{Rejection rates of $(h,\phi)$-divergence and Kolmogorov-Smirnov tests for scenarios (iii)-(v) for roughness pair $(\alpha_1,\alpha_2)=(-1.5,-5)$}                \label{table-31}                                                                                                                                                          \begin{tabular}{ r@{ }r@{ }c@{ }r@{ } r@{ } r@{ } r@{ } r@{ }c@{ }r@{ } r@{ } r@{ } r@{ } r@{ }c@{ }r@{ } r@{ } r@{ } r@{ } r@{ }}\toprule                                

\multicolumn{2}{c}{Constants}&\multicolumn{1}{c}{$\gamma$}& \multicolumn{5}{c}{$\mu_1>\mu_2$} &\multicolumn{1}{c}{$\gamma$}& \multicolumn{5}{c}{$\mu_1=\mu_2$}&\multicolumn{1}{c}{$\gamma$}&\multicolumn{5}{c}{$\mu_1<\mu_2$} \\

\cmidrule(lr{.25em}){1-2} \cmidrule(lr{.25em}){3-3} \cmidrule(lr{.25em}){4-8} \cmidrule(lr{.25em}){9-9} \cmidrule(lr{.25em}){10-14} \cmidrule(lr{.25em}){15-15} \cmidrule(lr{.25em}){16-20}                                     

$L$ & $N$ &$(\gamma_1,\gamma_2)$& $S_\text{KS}$ & $S_\text{KL}$ &  $S_\text{T}$ &  $S_\text{B}$&   $S_\text{AG}$ & $(\gamma_1,\gamma_2)$ & $S_\text{KS}$ &                $S_\text{KL}$ &  $S_\text{T}$ &  $S_\text{B}$&   $S_\text{AG}$ & $(\gamma_1,\gamma_2)$ & $S_\text{KS}$ & $S_\text{KL}$ &  $S_\text{T}$ &  $S_\text{B}$& $S_\text{AG}$\\   

\cmidrule(lr{.25em}){1-2} \cmidrule(lr{.25em}){3-3} \cmidrule(lr{.25em}){4-8} \cmidrule(lr{.25em}){9-9} \cmidrule(lr{.25em}){10-14} \cmidrule(lr{.25em}){15-15} \cmidrule(lr{.25em}){16-20}                                     

   1  & 49   &  (1,4)      &  0.65	   & 11.15   & 0.85	   & 4.07	  & 19.42	   & (0.5,4)  &   25.36   & 29.40   & 21.80   & 27.09   & 33.07   &    (0.5,8)    & 91.91  & 95.52    & 94.08    &  95.55	   & 95.96   \\           
                & 81   &             &  1.33	   & 25.82   & 3.85	   & 13.12	& 35.38	   &          &   54.31   & 60.22   & 53.02   & 57.79   & 63.30   &               & 99.76  & 99.92    & 99.92    &  99.92	   & 99.92   \\ 
                & 121  &             &  1.31	   & 44.16   & 15.45   & 32.35	& 53.23	   &          &   75.85   & 84.22   & 81.42   & 83.47   & 85.89   &               & 100.00 & 100.00   & 100.00   &  100.00	 & 100.00  \\ 

\cmidrule(lr{.25em}){1-2}  \cmidrule(lr{.25em}){4-8}  \cmidrule(lr{.25em}){10-14}  \cmidrule(lr{.25em}){16-20}                                                            

  8  & 49   &             &  3.02	   & 74.90   & 55.05   & 69.6	  & 80.11	   &          &   92.13   & 99.91   & 99.77   & 99.89   & 99.96   &               & 100.00 & 100.00   & 100.00   &  100.00	 & 100.00  \\             
     & 81   &             &  11.13   & 93.72   & 88.57   & 92.61	& 94.8	   &          &   99.95   & 100.00  & 100.00  & 100.00  & 100.00  &               & 100.00 & 100.00   & 100.00   &  100.00	 & 100.00  \\             
     & 121  &             &  20.38   & 97.87   & 97.43   & 97.76	& 98.03	   &          &   100.00  & 100.00  & 100.00  & 100.00  & 100.00  &               & 100.00 & 100.00   & 100.00   &  100.00	 & 100.00  \\             

\cmidrule(lr{.25em}){1-2} \cmidrule(lr{.25em}){3-3} \cmidrule(lr{.25em}){4-8} \cmidrule(lr{.25em}){9-9} \cmidrule(lr{.25em}){10-14} \cmidrule(lr{.25em}){15-15} \cmidrule(lr{.25em}){16-20}                                     

  1  & 49   &  (2.5,4)    &  60.47   & 90.83   & 78.27   & 87.63	& 93.55	   &  (1,8)   &   24.11   & 29.42   & 21.33   & 26.57   & 33.43   &    (0.5,20)   & 100.00 & 100.00   & 100.00   &  100.00	 & 100.00  \\             
     & 81   &             &  91.87   & 99.48   & 98.51   & 99.32	& 99.66	   &          &   53.78   & 59.83   & 53.71   & 58.03   & 62.63   &               & 100.00 & 100.00   & 100.00   &  100.00	 & 100.00  \\             
     & 121  &             &  98.84   & 100.00  & 100.00  & 100.00	& 100.00   &          &   75.71   & 84.85   & 81.74   & 84.20   & 86.11   &               & 100.00 & 100.00   & 100.00   &  100.00	 & 100.00  \\             

\cmidrule(lr{.25em}){1-2}  \cmidrule(lr{.25em}){4-8}  \cmidrule(lr{.25em}){10-14}  \cmidrule(lr{.25em}){16-20}                                                            

  8  & 49   &             &  99.49   & 100.00  & 99.98   & 100.00	& 100.00   &          &   92.13   & 99.89   & 99.79   & 99.89   & 99.91   &               & 100.00 & 100.00   & 100.00   &  100.00	 & 100.00  \\             
     & 81   &             &  99.98   & 100.00  & 100.00  & 100.00	& 100.00   &          &   99.80   & 100.00  & 100.00  & 100.00  & 100.00  &               & 100.00 & 100.00   & 100.00   &  100.00	 & 100.00  \\             
     & 121  &             &  100.00  & 100.00  & 100.00  & 100.00	& 100.00   &          &   100.00  & 100.00  & 100.00  & 100.00  & 100.00  &               & 100.00 & 100.00   & 100.00   &  100.00	 & 100.00  \\             

\cmidrule(lr{.25em}){1-2} \cmidrule(lr{.25em}){3-3} \cmidrule(lr{.25em}){4-8} \cmidrule(lr{.25em}){9-9} \cmidrule(lr{.25em}){10-14} \cmidrule(lr{.25em}){15-15} \cmidrule(lr{.25em}){16-20}                                     

  1  & 49   &  (2.5,8)    &  2.18	   & 21.98   & 4.46	   & 11.99	& 31.40	   & (2.5,20) &   23.85   & 27.70	  & 21.51   & 25.43   & 31.72   &    (1,20)     & 97.49  & 98.86    & 98.67    &  98.83	   & 99.02   \\             
     & 81   &             &  7.02	   & 45.56   & 15.82   & 33.52	& 55.15	   &          &   54.78   & 60.56   & 53.55   & 58.22   & 63.61   &               & 100.00 & 100.00   & 100.00   &  100.00	 & 100.00  \\             
     & 121  &             &  10.25   & 68.37   & 42.64   & 59.95	& 75.50	   &          &   75.38   & 85.19   & 81.98   & 84.22   & 86.58   &               & 100.00 & 100.00   & 100.00   &  100.00	 & 100.00  \\             

\cmidrule(lr{.25em}){1-2}  \cmidrule(lr{.25em}){4-8}  \cmidrule(lr{.25em}){10-14}  \cmidrule(lr{.25em}){16-20}                                                            

  8  & 49   &             &  13.67   & 79.52   & 58.63   & 74.64	& 84.73	   &          &   92.73   & 99.89   & 99.74   & 99.87   & 99.91   &               & 100.00 & 100.00   & 100.00   &  100.00	 & 100.00  \\             
     & 81   &             &  38.49   & 96.08   & 91.27   & 95.11	& 96.94	   &          &   99.89   & 100.00	& 100.00	& 100.00	& 100.00	&               & 100.00 & 100.00   & 100.00   &  100.00	 & 100.00  \\             
     & 121  &             &  62.85   & 99.69   & 99.05   & 99.6	  & 99.84	   &          &   100.00	& 100.00	& 100.00	& 100.00	& 100.00	&               & 100.00 & 100.00   & 100.00   &  100.00	 & 100.00  \\             

\bottomrule                                                                                                                                                               \end{tabular}                                                                                                                                                             \end{table*}                                                                                                                                                                                                                    

\begin{table*}[htb]                                                                                                                                                       \centering                                                                                                                                                                \scriptsize                                                                                                                                                               \caption{Rejection rates of $(h,\phi)$-divergence and Kolmogorov-Smirnov tests for scenarios (iii)-(v) for $(\alpha_1,\alpha_2)=(-3,-5)$}                                 \label{table-32}                                                                                                                                                          \begin{tabular}{ r@{ }r@{ }c@{ }r@{ } r@{ } r@{ } r@{ } r@{ }c@{ }r@{ } r@{ } r@{ } r@{ } r@{ }c@{ }r@{ } r@{ } r@{ } r@{ } r@{ }}\toprule                                \multicolumn{2}{c}{Constants}&\multicolumn{1}{c}{$\gamma$}& \multicolumn{5}{c}{$\mu_1>\mu_2$} &\multicolumn{1}{c}{$\gamma$}&                                              \multicolumn{5}{c}{$\mu_1=\mu_2$}&\multicolumn{1}{c}{$\gamma$}&\multicolumn{5}{c}{$\mu_1<\mu_2$} \\                                                                       \cmidrule(lr{.25em}){1-2} \cmidrule(lr{.25em}){3-3} \cmidrule(lr{.25em}){4-8} \cmidrule(lr{.25em}){9-9} \cmidrule(lr{.25em}){10-14} \cmidrule(lr{.25em}){15-15} \cmidrule(lr{.25em}){16-20}                         
$L$ & $N$ &$(\gamma_1,\gamma_2)$& $S_\text{KS}$ & $S_\text{KL}$ &  $S_\text{T}$ &  $S_\text{B}$&   $S_\text{AG}$ & $(\gamma_1,\gamma_2)$ & $S_\text{KS}$ &                $S_\text{KL}$ &  $S_\text{T}$ &  $S_\text{B}$&   $S_\text{AG}$ & $(\gamma_1,\gamma_2)$ & $S_\text{KS}$ & $S_\text{KL}$ &  $S_\text{T}$ &  $S_\text{B}$& $S_\text{AG}$\\   \cmidrule(lr{.25em}){1-2} \cmidrule(lr{.25em}){3-3} \cmidrule(lr{.25em}){4-8} \cmidrule(lr{.25em}){9-9} \cmidrule(lr{.25em}){10-14} \cmidrule(lr{.25em}){15-15} \cmidrule(lr{.25em}){16-20}                         
1  & 49   &  (4,4)      &  19.2	   & 32.15   & 20.93   & 28.45	& 37.20    &   (2,4)  &   1.07	  & 0.58	  & 0.39	  & 0.54	  & 1.00    &    (2,8)      & 46.45  & 52.51	  & 45.57	   &  51.89	   & 55.28   \\ 
& 81   &             &  44.62   & 62.18   & 51.78   & 59.56	& 65.80    &          &   1.96	  & 0.98	  & 0.58	  & 0.82	  & 1.49    &               & 81.96  & 83.39	& 80.52	   &  83.33	   & 84.91   \\ 
& 121  &             &  66.18   & 85.15   & 80.11   & 84.24	& 86.91    &          &   2.13	  & 1.52	  & 0.89	  & 1.12	  & 2.56    &               & 95.53  & 97.25	& 96.89	   &  97.22	   & 97.48   \\ 

\cmidrule(lr{.25em}){1-2}  \cmidrule(lr{.25em}){4-8}  \cmidrule(lr{.25em}){10-14}  \cmidrule(lr{.25em}){16-20}                                                            

  8  & 49   &             &  81.25   & 94.27   & 91.28   & 93.94	& 95.16    &          &   3.71	  & 12.14   & 6.36	  & 10.70	  & 15.60   &               & 99.62  & 99.92	  & 99.90    &  99.92	   & 99.96   \\ 
     & 81   &             &  98.49   & 98.75   & 98.6	   & 98.75	& 98.79    &          &   9.22	  & 26.17   & 19.21   & 24.48   & 29.42   &               & 100.00 & 100.00	  & 100.00	 &  100.00	 & 100.00  \\ 
     & 121  &             &  99.98   & 98.21   & 98.2	   & 98.21	& 98.21    &          &   14.55   & 42.30	  & 36.07   & 41.31   & 45.10   &               & 100.00 & 100.00	  & 100.00	 &  100.00	 & 100.00  \\ 

\cmidrule(lr{.25em}){1-2} \cmidrule(lr{.25em}){3-3} \cmidrule(lr{.25em}){4-8} \cmidrule(lr{.25em}){9-9} \cmidrule(lr{.25em}){10-14} \cmidrule(lr{.25em}){15-15} \cmidrule(lr{.25em}){16-20}                         

  1  & 49   &  (10,4)     &  98.71   & 99.81   & 99.61   & 99.77	& 99.81    &  (4,8)   &   1.16	  & 0.54	  & 0.38	  & 0.50	  & 1.15    &    (2,20)     & 99.78  & 100.00	  & 100.00	 &  100.00	 & 100.00  \\ 
     & 81   &             &  100.00	 & 100.00	 & 100.00	 & 100.00	& 100.00   &          &   1.69	  & 1.36	  & 0.81	  & 1.11	  & 1.90    &               & 100.00 & 100.00	  & 100.00	 &  100.00	 & 100.00  \\ 
     & 121  &             & 100.00	 & 100.00	 & 100.00	 & 100.00	& 100.00   &          &   2.53	  & 2.11	  & 1.27	  & 1.71	  & 2.93    &               & 100.00 & 100.00	  & 100.00	 &  100.00	 & 100.00  \\ 

\cmidrule(lr{.25em}){1-2}  \cmidrule(lr{.25em}){4-8}  \cmidrule(lr{.25em}){10-14}  \cmidrule(lr{.25em}){16-20}                                                            

  8  & 49   &             & 100.00	 & 100.00	 & 100.00	 & 100.00	& 100.00   &          &   3.15	  & 13.05   & 6.36	  & 11.37   & 16.25   &               & 100.00 & 100.00	  & 100.00	 &  100.00	 & 100.00  \\ 
     & 81   &             & 100.00	 & 100.00	 & 100.00	 & 100.00	& 100.00   &          &   8.56	  & 25.55   & 18.48   & 23.98   & 28.87   &               & 100.00 & 100.00	  & 100.00	 &  100.00	 & 100.00  \\ 
     & 121  &             & 100.00	 & 100.00	 & 100.00	 & 100.00	& 100.00   &          &   12.98   & 43.64   & 36.85   & 42.35   & 46.68   &               & 100.00 & 100.00	  & 100.00	 &  100.00	 & 100.00  \\ 

\cmidrule(lr{.25em}){1-2} \cmidrule(lr{.25em}){3-3} \cmidrule(lr{.25em}){4-8} \cmidrule(lr{.25em}){9-9} \cmidrule(lr{.25em}){10-14} \cmidrule(lr{.25em}){15-15} \cmidrule(lr{.25em}){16-20}                         

  1  & 49   &  (10,8)     & 46.93    & 65.05   & 52.45   & 62.71	& 69.91    &  (10,20) &   0.91	  & 0.70	  & 0.31	  & 0.54	  & 1.05    &     (4,20)    & 74.84  & 81.73	  & 76.48	   &  81.42	   & 84.18   \\ 
     & 81   &             &  81.65   & 91.53   & 87.37   & 90.84	& 92.98    &          &   2.18	  & 1.30	  & 0.96	  & 1.06	  & 2.11    &               & 96.71  & 98.04	  & 97.77	   &  98.07	   & 98.25   \\ 
     & 121  &             &  94.95   & 99.06   & 98.55   & 98.88	& 99.36    &          &   2.29	  & 2.19	  & 1.25	  & 1.58	  & 3.10    &               & 99.71  & 99.95	  & 99.92	   &  99.95	   & 99.95   \\ 

\cmidrule(lr{.25em}){1-2}  \cmidrule(lr{.25em}){4-8}  \cmidrule(lr{.25em}){10-14}  \cmidrule(lr{.25em}){16-20}                                                            

  8  & 49   &             &  99.22   & 99.98   & 99.94   & 99.98	& 99.98    &          &   2.96	  & 12.71   & 6.67	  & 11.31   & 16.28   &               & 100.00 & 100.00	  & 100.00	 &  100.00	 & 100.00   \\
     & 81   &             &  100.00	 & 100.00	 & 100.00	 & 100.00	& 100.00   &          &   8.91	  & 26.29   & 19.33   & 24.93   & 29.79   &               & 100.00 & 100.00	  & 100.00	 &  100.00	 & 100.00   \\
     & 121  &             & 100.00	 & 100.00	 & 100.00	 & 100.00	& 100.00   &          &   13.58   & 42.24   & 36.07   & 41.11   & 44.92   &               & 100.00 & 100.00	  & 100.00	 &  100.00	 & 100.00   \\

\bottomrule                                                                                                                                                               
\end{tabular}                                                                                                                                                             \end{table*}                                                                                                                                                              

\subsection{ML Estimation in Contaminated Data}\label{sec:simul1}

In practice, it is quite hard to guarantee that all observations in a sample come from random variables with the same distribution.
When collecting samples by visual inspection,
frequently the onset of some type of contamination is experienced.

\begin{figure}[h]
\centering
\includegraphics[width=.5\linewidth]{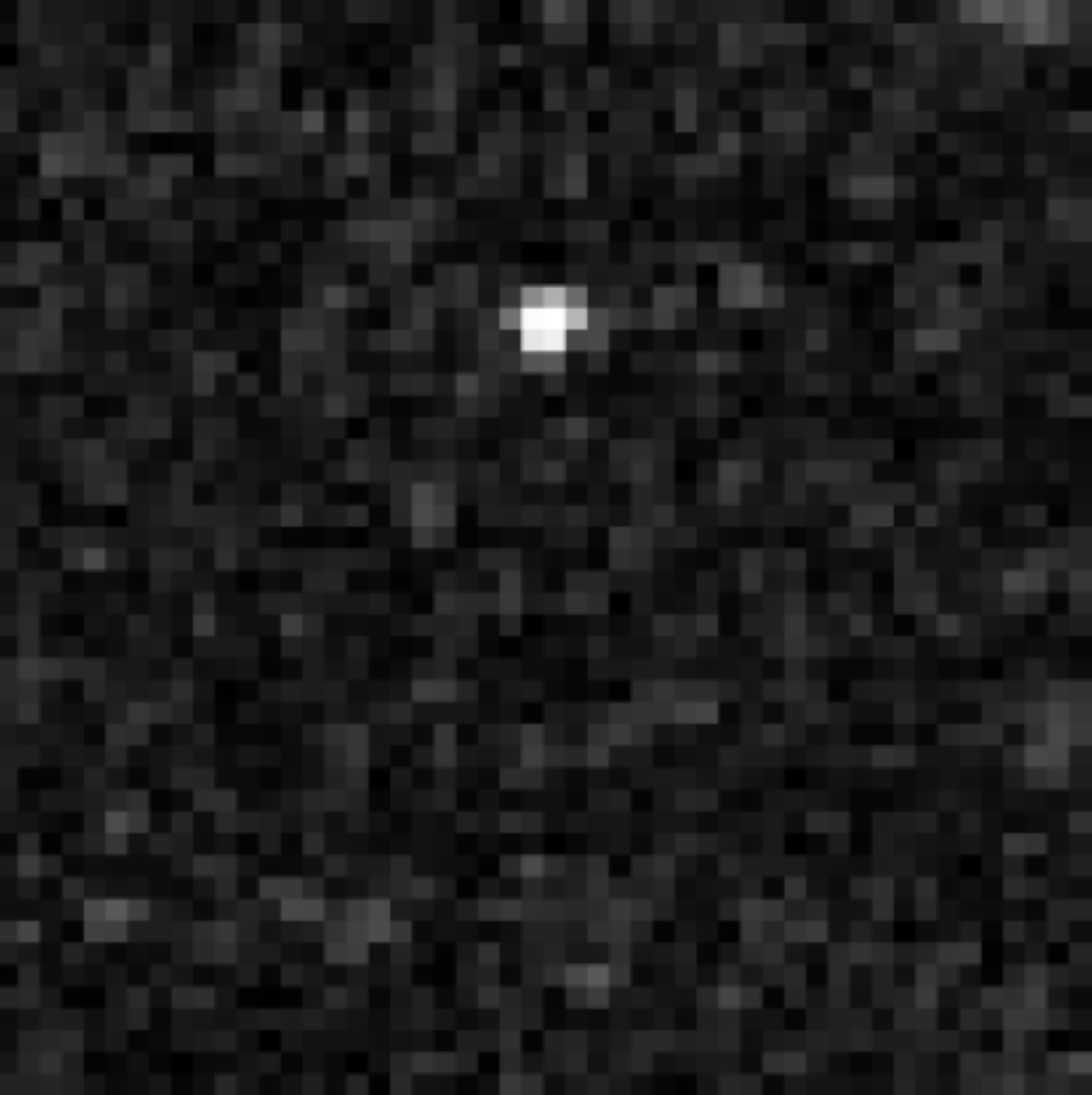}%
\caption{A corner reflector producing double bounce on a homogeneous background}\label{fig:DoubleBounce}
\end{figure}

One of the most common sources of contamination in SAR imagery is related to the \textit{double bounce} phenomenon: a few pixels exhibit a very high return caused by either natural or man-made targets that send back most of the energy they receive.
This is the case of flooded forests and urban areas.
The presence of a single pixel suffering from double bounce may produce meaningless estimates and test statistics, unless they exhibit some robustness with respect to
this kind of contamination.
Fig.~\ref{fig:DoubleBounce} shows an example of the double bounce effect on an homogeneous field.

In order to assess the robustness of the considered tests, we assume contamination in a parametric way: the data in a sample come either from the hypothesized $\mathcal{G}^0$ distribution with probability $1-\epsilon$, or from a scaled version of the same law with probability $\epsilon$.
The considered scaled distribution is such that its mean value is one hundred times larger than the mean of the original distribution.
Let $A$ be the event ``presence of outlier", which has $\Pr{(A)}=\epsilon$, and $A'$ its complement.
By the total probability rule, the cumulative distribution function of the return is
\begin{align*}\label{contamin}
F_{Z}(z)&=\Pr{(Z\leq z)}\nonumber\\
&=\Pr{\bigl(\{Z\leq z\}\cap A'\bigr)}+\Pr{\bigl(\{Z\leq z\}\cap A\bigr)} \nonumber \\
&=\Pr{(A')\Pr{(Z\leq z\mid A')}}+\Pr{(A)}\Pr{(Z\leq z\mid A)} \nonumber \\
&=(1-\epsilon)F_{\mathcal{G}^0(\alpha,\gamma,L)}(z)+ \epsilon F_{\mathcal{G}^0(\alpha,100\gamma,L)}(z).
\end{align*}
Following~\cite{AllendeFreryetal:JSCS:05,BustosFreryLucini:Mestimators:2001}, we chose to assess the situations without contamination $\epsilon=0$ and with mild levels of contamination $\epsilon\in \{10^{-4}, 5\times 10^{-3}\}$.

Simulation results are based on $5500$ independent replications, and instances were considered valid only if $\widehat\alpha\in[10\alpha,\alpha/20]$, i.e., censoring was applied as in~\cite{FreryandCribariNetoandSouza2004,HypothesisTestingSpeckledDataStochasticDistances}.
Ta\-bles~\ref{tableestimationML},~\ref{tableestimationML1}, and~\ref{tableestimationML2} present the estimated bias (B), the coefficient of variation (CV, in percentage), and the mean squared error (MSE) of the ML estimators $\widehat{\alpha}$ and $\widehat{\gamma}$ for the roughness degrees $-1.5$, $-3$, and~$-5$ according to several simulation conditions and contamination levels, respectively.

Let us first consider the bias for uncontaminated data.
The only situations for which the bias reduces with in\-crea\-sing sample size are: (a)~$\alpha=-1.5$, (b)~$\alpha=-3$, $L=8$, and (c)~$\alpha=-5$, $L=8$.
This seemingly surprising situation is frequent when dealing with the $\mathcal G^0$ distribution.
Indeed, the information content is reduced in homogeneous and heterogeneous situations; therefore, the good asymptotic properties of ML estimators are not verified in samples of the considered size~\cite{CribariFrerySilva:CSDA,FreryandCribariNetoandSouza2004,VasconcellosandFreryandSilva2005}.
An increase of the number of looks can partially compensate this behavior.

This loss of information content is also verified by the dependence of the bias on the number of looks.
It is expected that the former is reduced increasing the latter, but this is only verified when $\alpha=-1.5$, i.e., in extremely heterogeneous areas.
When dealing with heterogeneous data ($\alpha=-3$), this reduction only takes place starting at $N=81$, and with homogeneous data ($\alpha=-5$) at $N=121$.
Again, the more homogeneous the sample is, the smaller its information content becomes under the $\mathcal G^0$ model.
Regarding the $N=121$ and $L=8$ situation, which is expected to reveal the asymptotic behavior of the estimators, the bias increases when $\alpha$ is reduced from $-1.5$ to $-5$.

Regarding the influence of contamination on the bias, in all but a single case (namely $N=49, L=1, \epsilon=10^{-4}$), increasing the former increases the latter for both estimators.
This increase in the bias ranges from, approximately, $34\%$ ($\alpha=-1.5,L=1,\gamma=0.5,\epsilon=10^{-4}$) to $1000$\% ($\alpha=-3,L=1,\gamma=2,\epsilon=5\times 10^{-3}$), in both cases $N=49$.

The coefficient of variation is reduced by increasing $L$ and/or $N$ and/or $\alpha$, as expected, in uncontaminated data.
In the presence of contamination, this behavior is lost with res\-pect to $\alpha$.

The usual behavior of reduced MSE with increasing $L$ and/or $N$ and/or $\alpha$ is observed in the absence of contamination.
The presence of contamination consistently increases the MSE, except in four out of $324$ situations, where random fluctuations masked the results.
For instance, the MSE of $\widehat\gamma$ doubled in the $\alpha=-5$, $\gamma=20$, $L=1$, $N=49$ situation when $\epsilon=10^{-4}$, i.e., the mildest level of contamination.
If $\alpha=-5$, $\gamma=20$, $L=8$, $N=121$, and $\epsilon=5\times10^{-3}$, i.e., the strongest contamination, the MSE was affected by a factor of $13794$.

This strong influence of contamination on the MSE warns users about the need of quality samples and of robust estimators when dealing with this type of data.

\begin{table*}[htb]
\centering
\scriptsize
\caption{Maximum likelihood estimates bias (B), coefficients of variation (CV\%), and mean squared errors (MSE) for high roughness}
\label{tableestimationML}
\begin{tabular}{@{ }c@{ }c@{ }c  c@{ }  r@{  }r@{ }r r@{ }r@{ }r r@{ }r@{ }r r@{ }r@{ }r r@{ }r@{ }r r@{ }r@{ }r@{ }}\toprule
&&&  &\multicolumn{9}{c}{$\widehat\alpha$} &\multicolumn{9}{c}{$\widehat\gamma$}\\ \cmidrule(lr{.25em}){5-13} \cmidrule(lr{.25em}){14-22}
$\alpha$ & $\gamma$ & $L$ & $\epsilon$ &\multicolumn{3}{c}{$N=49$} &\multicolumn{3}{c}{$N=81$} &\multicolumn{3}{c}{$N=121$} &\multicolumn{3}{c}{$N=49$} &\multicolumn{3}{c}{$N=81$} &\multicolumn{3}{c}{$N=121$} \\
&&& & \multicolumn{1}{c}{B} & {CV\%} & {MSE}
& \multicolumn{1}{c}{B} & {CV\%} & {MSE} 
& \multicolumn{1}{c}{B} & {CV\%} & {MSE} 
& \multicolumn{1}{c}{B} & {CV\%} & {MSE} 
& \multicolumn{1}{c}{B} & {CV\%} & {MSE} 
& \multicolumn{1}{c}{B} & {CV\%} & {MSE} \\ \cmidrule(lr{.25em}){1-3} \cmidrule(lr{.25em}){4-4} \cmidrule(lr{.25em}){5-13} \cmidrule(lr{.25em}){14-22}
                                                                                                                            
$-1.5$ & $0.5$  & $1$ & $0$      &  0.23    & \ 37.64	  & 0.47	& 0.18   & \ 32.50	 & 0.33	 & 0.14 & \ 27.15	 & 0.22	 &   0.12  & \ 56.10	 & 0.13	    &  0.09 & \ 47.64	 &  0.09	  &  0.07  &  \ 40.00	& 0.06   \\                              
                                                            
    &      & $8$     &           &  0.11    & 23.58	  & 0.15	& 0.06   & 17.39	 & 0.08	 & 0.04 & 13.95	 & 0.05	 &   0.05  &  29.83	 & 0.03	    &  0.03 & 21.76	 &  0.01	  &  0.02  &  17.67	& 0.01   \\                              
                                                            
    &   $1$  & $1$   &           &  0.22    & 37.70	  & 0.50	& 0.17   & 32.24	 & 0.34	 & 0.12 & 26.86	 & 0.21	 &   0.24  &  56.60	 & 0.60	    &  0.19 & 47.66	 &  0.37	  &  0.13  &  39.59	& 0.23   \\                              
                                                            
     &      & $8$     &          &  0.10    & 22.73	  & 0.15	& 0.06   & 17.30	 & 0.08	 & 0.05 & 14.01	 & 0.05	 &   0.09  &  28.24	 & 0.11	    &  0.05 & 21.78	 &  0.06	  &  0.04  &  17.64	& 0.04   \\                              
                                                            
     & $2.5$  & $1$   &          &  0.20    & 36.63	  & 0.47	& 0.19   & 32.59	 & 0.33	 & 0.13 & 26.74	 & 0.20	 &   0.55  &  54.94	 & 3.47	    &  0.49 & 48.49	 &  2.25	  &  0.35  &  39.55	& 1.35   \\                              
                                                            
     &      & $8$     &          &  0.12    & 23.31	  & 0.15	& 0.06   & 17.38	 & 0.08	 & 0.04 & 13.87	 & 0.05	 &   0.26  &  29.26	 & 0.66	    &  0.14 & 21.88	 &  0.35	  &  0.09  &  17.45	& 0.22   \\  \cmidrule(lr{.25em}){4-4} \cmidrule(lr{.25em}){5-13} \cmidrule(lr{.25em}){14-22}
                                                            
     &      &         &$10^{-4}$ &  0.31	  & 36.75	  & 0.54	& 0.27	 & 32.33	 & 0.40	 & 0.23	& 27.15	 & 0.27	 &   0.19  &  52.79	 & 0.17	    &  0.16 & 46.73	 &  0.12	  &  0.14  &  39.15	& 0.08   \\                              
                                                            
     &      &         &          &  0.17	  & 23.78	  & 0.19	& 0.12	 & 18.03	 & 0.10	 & 0.10	& 14.08	 & 0.06	 &   0.09  &  28.50	 & 0.04	    &  0.07 & 21.98	 &  0.02	  &  0.06  &  17.14	& 0.01   \\                              
                                                            
     &      &         &          &  0.29	  & 36.41	  & 0.51	& 0.26	 & 31.36	 & 0.37	 & 0.23	& 27.19	 & 0.28	 &   0.37  &  53.47	 & 0.67	    &  0.31 & 45.41	 &  0.45	  &  0.28  &  38.05	& 0.31   \\                              
                                                            
     &      &         &          &  0.17	  & 23.51	  & 0.18	& 0.13	 & 17.45	 & 0.10	 & 0.10	& 14.16	 & 0.06	 &   0.18  &  28.72	 & 0.15	    &  0.14 & 20.98	 &  0.08	  &  0.12  &  17.53	& 0.05   \\                              
                                                            
     &      &         &          &  0.30	  & 36.71	  & 0.52	& 0.28	 & 32.05	 & 0.40	 & 0.23	& 26.78	 & 0.26	 &   0.91  &  52.60	 & 4.05	    &  0.83 & 45.77	 &  3.01	  &  0.67  &  38.11	& 1.91   \\                              
                                                            
     &      &         &          &  0.17	  & 23.34	  & 0.18	& 0.12	 & 17.54	 & 0.09	 & 0.10	& 14.20	 & 0.06	 &   0.47  &  28.79	 & 0.95	    &  0.34 & 21.54	 &  0.49	  &  0.31  &  17.48	& 0.33   \\ \cmidrule(lr{.25em}){4-4} \cmidrule(lr{.25em}){5-13} \cmidrule(lr{.25em}){14-22}
                                                            
     &      &         &$10^{-3}$ &  0.94	  & 47.42	  & 2.22	& 0.84	 & 42.28	 & 1.68	 & 0.75	& 35.29	 & 1.19	 &   0.74  &  60.30	 & 1.10	    &  0.66 & 53.66	 &  0.82	  &  0.60  &  44.81	& 0.60   \\                              
                                                            
     &      &         &          &  0.44	  & 25.90	  & 0.44	& 0.38	 & 19.51	 & 0.28	 & 0.35	& 15.16	 & 0.20	 &   0.35  &  29.15	 & 0.19	    &  0.32 & 22.41	 &  0.14	  &  0.30  &  17.38	& 0.11   \\                              
                                                            
     &      &         &          &  0.91	  & 47.46	  & 2.14	& 0.82	 & 40.98	 & 1.58	 & 0.75	& 34.74	 & 1.17	 &   1.44  &  60.88	 & 4.30	    &  1.31 & 52.46	 &  3.17	  &  1.20  &  43.15	& 2.33   \\                              
                                                            
     &      &         &          &  0.44	  & 25.98	  & 0.44	& 0.38	 & 19.06	 & 0.27	 & 0.35	& 15.14	 & 0.20	 &   0.70  &  29.95	 & 0.75	    &  0.64 & 21.67	 &  0.53	  &  0.61  &  17.52	& 0.45   \\                              
                                                            
     &      &         &          &  0.93	  & 48.01	  & 2.23	& 0.85	 & 41.88	 & 1.69	 & 0.75	& 34.69	 & 1.16	 &   3.65  &  60.33	 & 27.04	  &  3.34 & 52.76	 &  20.68	  &  2.98  &  43.82	& 14.68  \\                              
                                                            
     &      &         &          &  0.44	  & 25.83	  & 0.45	& 0.37	 & 18.84	 & 0.26	 & 0.35	& 15.12	 & 0.20	 &   1.76  &  29.81	 & 4.73	    &  1.58 & 21.76	 &  3.28	  &  1.52  &  17.47	& 2.81   \\ \cmidrule(lr{.25em}){4-4} \cmidrule(lr{.25em}){5-13} \cmidrule(lr{.25em}){14-22}
                                                            
     &      &  &$5\times 10^{-3}$&  1.45	  & 44.24	  & 3.79	& 1.42	 & 40.45	 & 3.42	 & 1.41	& 36.47	 & 3.11	 &   1.93  &  53.58	 & 5.41	    &  1.87 & 48.52	 &  4.82	  &  1.85  &  43.66	& 4.45   \\                              
                                                            
     &      &         &          &  0.75	  & 26.83	  & 0.93	& 0.69	 & 20.99	 & 0.68	 & 0.64	& 16.35	 & 0.53	 &   1.05  &  29.32	 & 1.30	    &  1.00 & 23.29	 &  1.11	  &  0.96  &  18.29	& 1.00   \\                              
                                                            
     &      &         &          &  1.43	  & 44.82	  & 3.78	& 1.43	 & 39.83	 & 3.40	 & 1.42	& 36.66	 & 3.16	 &   3.84  &  55.04	 & 21.86	  &  3.76 & 47.85	 &  19.30	  &  3.71  &  43.21	& 17.90  \\                              
                                                            
     &      &         &          &  0.75	  & 27.64	  & 0.95	& 0.68	 & 20.33	 & 0.66	 & 0.64	& 16.13	 & 0.53	 &   2.10  &  30.80	 & 5.31	    &  1.98 & 22.65	 &  4.37	  &  1.92  &  18.05	& 3.96   \\                              
                                                            
     &      &         &          &  1.46	  & 45.50	  & 3.93	& 1.42	 & 40.18	 & 3.40	 & 1.41	& 36.94	 & 3.16	 &   9.69  &  55.03	 & 138.98	  &  9.36	 & 48.42	 &  120.59  &  9.24  &  43.82	& 111.92 \\                            
                                                            
     &      &         &          &  0.76	  & 27.45	  & 0.97	& 0.68	 & 20.17	 & 0.65	 & 0.64	& 16.01	 & 0.53	 &   5.28  &  30.76	 & 33.60	  &  4.95 & 22.63	 &  27.39	  &  4.82  &  17.80	& 24.89  \\                       
\bottomrule
\end{tabular}                                                                                                                  
\end{table*}
                                                            
\begin{table*}[htb]
\centering
\scriptsize
\caption{Maximum likelihood estimates bias (B), coefficients of variation (CV\%), and mean squared errors (MSE) for mid roughness}
\label{tableestimationML1}
\begin{tabular}{@{ }c@{ }c@{ }c  c@{ }  r@{  }r@{ }r r@{ }r@{ }r r@{ }r@{ }r r@{ }r@{ }r r@{ }r@{ }r r@{ }r@{ }r@{ }}\toprule
&&&  &\multicolumn{9}{c}{$\widehat\alpha$} &\multicolumn{9}{c}{$\widehat\gamma$}\\ \cmidrule(lr{.25em}){5-13} \cmidrule(lr{.25em}){14-22}
$\alpha$ & $\gamma$ & $L$ & $\epsilon$ &\multicolumn{3}{c}{$N=49$} &\multicolumn{3}{c}{$N=81$} &\multicolumn{3}{c}{$N=121$} &\multicolumn{3}{c}{$N=49$} &\multicolumn{3}{c}{$N=81$} &\multicolumn{3}{c}{$N=121$} \\
&&& & \multicolumn{1}{c}{B} & {CV\%} & {MSE}
& \multicolumn{1}{c}{B} & {CV\%} & {MSE} 
& \multicolumn{1}{c}{B} & {CV\%} & {MSE} 
& \multicolumn{1}{c}{B} & {CV\%} & {MSE} 
& \multicolumn{1}{c}{B} & {CV\%} & {MSE} 
& \multicolumn{1}{c}{B} & {CV\%} & {MSE} \\ \cmidrule(lr{.25em}){1-3} \cmidrule(lr{.25em}){4-4} \cmidrule(lr{.25em}){5-13} \cmidrule(lr{.25em}){14-22}

$-3.0$ &  $2.0$ & $1$ & $0$      &  0.15	  & 42.11	  & 1.81	& 0.29	 & 37.21	 & 1.62	 & 0.29	& 34.61	 & 1.42	 &   0.16  &  56.46	 & 1.55	    &  0.25 & 48.36	 &  1.27	  &  0.25  &  44.09	& 1.12   \\                              
                                                            
     &      & $8$     &          &  0.31	  & 27.28	  & 0.90	& 0.18	 & 22.14	 & 0.49	 & 0.12	& 17.58	 & 0.30	 &   0.24  &  31.38	 & 0.55	    &  0.14 & 25.21	 &  0.29	  &  0.09  &  20.19	& 0.18   \\                              
                                                            
     &  $4.0$ & $1$   &          &  0.17	  & 41.35	  & 1.79	& 0.29	 & 37.47	 & 1.63	 & 0.30	& 34.99	 & 1.45	 &   0.35  &  55.24	 & 5.95	    &  0.53 & 48.57	 &  5.35	  &  0.53  &  44.84	& 4.33   \\                              
                                                            
     &      & $8$     &          &  0.29	  & 27.51	  & 0.88	& 0.18	 & 21.29	 & 0.49	 & 0.11	& 18.28	 & 0.32	 &   0.45  &  31.65	 & 2.14	    &  0.28 & 24.47	 &  1.17	  &  0.18  &  20.84	& 0.75   \\                              
                                                            
     & $10.0$ & $1$   &          &  0.18	  & 41.92	  & 1.84	& 0.32	 & 38.17	 & 1.66	 & 0.31	& 35.07	 & 1.51	 &   0.90  &  55.33	 & 39.62	  &  1.42	 & 49.42	 &  32.20	  &  1.38  &  44.93	& 29.69  \\                            
                                                            
     &      & $8$     &          &  0.28	  & 27.79	  & 0.88	& 0.16	 & 21.15	 & 0.51	 & 0.12	& 17.93	 & 0.29	 &   1.09  &  31.91	 & 13.19	  &  0.65	 & 24.30	 &  7.71	  &  0.48  &  20.61	& 4.36   \\ \cmidrule(lr{.25em}){4-4} \cmidrule(lr{.25em}){5-13} \cmidrule(lr{.25em}){14-22}
                                                            
     &      &         &$10^{-4}$ &  0.32	  & 41.92	  & 2.03	& 0.43	 & 37.48	 & 1.84	 & 0.46	& 34.04	 & 1.60	 &   0.33  &  54.06	 & 1.69	    &  0.42 & 48.17	 &  1.54	  &  0.44  &  43.13	& 1.31   \\                              
                                                            
     &      &         &          &  0.40	  & 27.47	  & 1.03	& 0.29	 & 21.30	 & 0.57	 & 0.23	& 17.48	 & 0.37	 &   0.35  &  31.15	 & 0.66	    &  0.26 & 24.49	 &  0.37	  &  0.22  &  19.85	& 0.24   \\                              
                                                            
     &      &         &          &  0.33	  & 40.13	  & 1.90	& 0.47	 & 37.01	 & 1.87	 & 0.47	& 33.76	 & 1.59	 &   0.75  &  52.75	 & 6.85	    &  0.92 & 47.45	 &  6.29	  &  0.91  &  43.08	& 5.30   \\                              
                                                            
     &      &         &          &  0.38	  & 27.10	  & 0.99	& 0.29	 & 21.72	 & 0.59	 & 0.23	& 17.18	 & 0.36	 &   0.68  &  30.78	 & 2.53	    &  0.53 & 24.77	 &  1.53	  &  0.43  &  19.70	& 0.94   \\                              
                                                            
     &      &         &          &  0.31	  & 41.12	  & 1.95	& 0.40	 & 36.90	 & 1.74	 & 0.48	& 33.94	 & 1.63	 &   1.68  &  54.01	 & 42.65	  &  2.05	 & 47.10	 &  36.42	  &  2.31  &  42.67	& 32.91  \\                            
                                                            
     &      &         &          &  0.39	  & 27.33	  & 1.01	& 0.30	 & 21.61	 & 0.60	 & 0.23	& 17.03	 & 0.36	 &   1.68  &  30.98	 & 15.93	  &  1.39	 & 24.67	 &  9.82	  &  1.09  &  19.26	& 5.74   \\ \cmidrule(lr{.25em}){4-4} \cmidrule(lr{.25em}){5-13} \cmidrule(lr{.25em}){14-22}
                                                            
     &      &         &$10^{-3}$ &  0.84	  & 35.35	  & 2.54	& 1.14	 & 31.11	 & 2.95	 & 1.30	& 28.34	 & 3.18	 &   1.22  &  44.44	 & 3.55	    &  1.49 & 38.04	 &  3.98	  &  1.65  &  34.06	& 4.28   \\                              
                                                            
     &      &         &          &  1.07	  & 25.91	  & 2.24	& 1.03	 & 21.99	 & 1.84	 & 0.97	& 18.74	 & 1.50	 &   1.27  &  28.36	 & 2.49	    &  1.24 & 24.24	 &  2.15	  &  1.18  &  20.27	& 1.82   \\                              
                                                            
     &      &         &          &  0.93	  & 34.41	  & 2.68	& 1.16	 & 30.90	 & 3.01	 & 1.34	& 27.64	 & 3.22	 &   2.70  &  43.67	 & 15.85	  &  3.04 & 37.61	 &  16.23	  &  3.35  &  33.24	& 17.21  \\                              
                                                            
     &      &         &          &  1.07	  & 25.88	  & 2.25	& 1.03	 & 22.05	 & 1.84	 & 0.97	& 18.47	 & 1.48	 &   2.55  &  28.48	 & 10.00	  &  2.47 & 24.07	 &  8.53	  &  2.37  &  20.11	& 7.26   \\                              
                                                            
     &      &         &          &  0.87	  & 35.12	  & 2.59	& 1.10	 & 31.03	 & 2.84	 & 1.35	& 28.24	 & 3.33	 &   6.31  &  44.07	 & 91.51	  &  7.47	 & 38.16	 &  100.21  &  8.47  &  33.70	& 110.56 \\                            
                                                            
     &      &         &          &  1.07	  & 25.95	  & 2.26	& 1.05	 & 22.39	 & 1.93	 & 0.99	& 18.49	 & 1.52	 &   6.35  &  28.37	 & 61.86	  &  6.31	 & 24.51	 &  55.79	  &  5.99  &  19.97	& 46.09  \\ \cmidrule(lr{.25em}){4-4} \cmidrule(lr{.25em}){5-13} \cmidrule(lr{.25em}){14-22}
                                                            
     &      &  &$5\times 10^{-3}$&  1.64	  & 27.29	  & 4.29	& 1.93	 & 24.11	 & 5.12	 & 2.19	& 20.16	 & 5.91	 &   3.99  &  33.89	 & 20.00	  &  4.33 & 28.75	 &  22.08	  &  4.72  &  23.54	& 24.78  \\                              
                                                            
     &      &         &          &  2.03	  & 20.87	  & 5.22	& 2.18	 & 18.18	 & 5.62	 & 2.28	& 15.81	 & 5.87	 &   4.05  &  23.00	 & 18.33	  &  4.22 & 19.88	 &  19.32	  &  4.34  &  17.00	& 19.96  \\                              
                                                            
     &      &         &          &  1.62	  & 28.30	  & 4.34	& 1.96	 & 23.49	 & 5.18	 & 2.26	& 19.92	 & 6.20	 &   7.85  &  34.20	 & 77.99	  &  8.79	 & 28.54	 &  90.59	  &  9.61  &  23.78	& 102.92 \\                            
                                                            
     &      &         &          &  2.03	  & 20.69	  & 5.23	& 2.18	 & 17.96	 & 5.64	 & 2.27	& 15.97	 & 5.88	 &   8.10  &  22.78	 & 73.24	  &  8.47	 & 19.56	 &  77.78	  &  8.67  &  17.21	& 79.98  \\                            
                                                            
     &      &         &          &  1.52	  & 28.95	  & 4.02	& 1.88	 & 23.62	 & 4.86	 & 2.14	& 21.23	 & 5.76	 &  19.13  &  35.19	 & 471.04	  & 21.52	 & 27.91	 &  540.58  & 23.29  &  25.41	& 614.05 \\                            
                                                            
     &      &         &          &  2.05	  & 20.42	  & 5.27	& 2.19	 & 18.12	 & 5.69	 & 2.30	& 15.95	 & 6.00	 &  20.35  &  22.38	 & 460.41	  & 21.22	 & 19.70	 &  488.21  & 21.85  &  17.19	& 507.52 \\                     
\bottomrule
\end{tabular}                                                                                                                  
\end{table*}

\begin{table*}[htb]
\centering
\scriptsize
\caption{Maximum likelihood estimates bias (B), coefficients of variation (CV\%), and mean squared errors (MSE) for  low roughness}
\label{tableestimationML2}
\begin{tabular}{@{ }c@{ }c@{ }c  c@{ }  r@{  }r@{ }r r@{ }r@{ }r r@{ }r@{ }r r@{ }r@{ }r r@{ }r@{ }r r@{ }r@{ }r@{ }}\toprule
&&&  &\multicolumn{9}{c}{$\widehat\alpha$} &\multicolumn{9}{c}{$\widehat\gamma$}\\ \cmidrule(lr{.25em}){5-13} \cmidrule(lr{.25em}){14-22}
$\alpha$ & $\gamma$ & $L$ & $\epsilon$ &\multicolumn{3}{c}{$N=49$} &\multicolumn{3}{c}{$N=81$} &\multicolumn{3}{c}{$N=121$} &\multicolumn{3}{c}{$N=49$} &\multicolumn{3}{c}{$N=81$} &\multicolumn{3}{c}{$N=121$} \\
&&& & \multicolumn{1}{c}{B} & {CV\%} & {MSE}
& \multicolumn{1}{c}{B} & {CV\%} & {MSE} 
& \multicolumn{1}{c}{B} & {CV\%} & {MSE} 
& \multicolumn{1}{c}{B} & {CV\%} & {MSE} 
& \multicolumn{1}{c}{B} & {CV\%} & {MSE} 
& \multicolumn{1}{c}{B} & {CV\%} & {MSE} \\ \cmidrule(lr{.25em}){1-3} \cmidrule(lr{.25em}){4-4} \cmidrule(lr{.25em}){5-13} \cmidrule(lr{.25em}){14-22}

$-5.0$ &  $4.0$ & $1$ & $0$      &  0.33	  & 46.97	  & 4.85	& 0.02	 & 42.92	 & 4.75	 & 0.22	& 39.77	 & 4.27	 &   0.30  &  58.61	 & 4.73	    &  0.03 & 51.94	 &  4.58	  &  0.22  &  47.82	& 3.90   \\                              
                                                            
     &      & $8$     &          &  0.52	  & 30.38	  & 3.01	& 0.35	 & 28.62	 & 2.04	 & 0.13	& 26.71	 & 1.28	 &   0.47  &  33.62	 & 2.42	    &  0.33 & 31.08	 &  1.61	  &  0.13  &  28.51	& 1.02   \\                              
                                                            
     &  $8.0$ & $1$   &          &  0.30	  & 46.86	  & 5.11	& 0.06	 & 42.50	 & 4.75	 & 0.32	& 40.23	 & 4.23	 &   0.48  &  57.82	 & 19.89	  &  0.05 & 51.74	 &  17.98	  &  0.64  &  48.68	& 15.97  \\                              
                                                            
     &      & $8$     &          &  0.51	  & 30.85	  & 3.14	& 0.28	 & 29.04	 & 2.08	 & 0.11	& 27.77	 & 1.28	 &   0.92  &  34.18	 & 10.23	  &  0.52 & 31.38	 &  6.64	  &  0.21  &  29.41	& 4.06   \\                              
                                                            
     & $20.0$ & $1$   &          &  0.25	  & 47.26	  & 4.99	& 0.00	 & 43.57	 & 4.74	 & 0.18	& 39.53	 & 4.13	 &   1.09  &  58.27	 & 122.69	  &  0.05	 & 53.04	 &  110.26  &  0.89  &  47.03	& 98.05  \\                            
                                                            
     &      & $8$     &          &  0.52	  & 30.61	  & 2.99	& 0.31	 & 28.25	 & 2.04	 & 0.13	& 27.63	 & 1.30	 &   2.28  &  33.64	 & 60.17	  &  1.43	 & 30.69	 &  39.68	  &  0.63  &  29.41	& 25.36  \\ \cmidrule(lr{.25em}){4-4} \cmidrule(lr{.25em}){5-13} \cmidrule(lr{.25em}){14-22}
                                                            
     &      &         &$10^{-4}$ &  0.14	  & 46.39	  & 5.11	& 0.20	 & 42.56	 & 4.93	 & 0.39	& 37.67	 & 4.26	 &   0.02  &  57.34	 & 5.21	    &  0.27 & 51.54	 &  4.93	  &  0.45  &  44.61	& 4.15   \\                              
                                                            
     &      &         &          &  0.70	  & 30.21	  & 3.44	& 0.57	 & 25.28	 & 2.31	 & 0.46	& 21.34	 & 1.57	 &   0.68  &  33.27	 & 2.90	    &  0.57 & 27.83	 &  1.95	  &  0.47  &  23.43	& 1.32   \\                              
                                                            
     &      &         &          &  0.14	  & 46.19	  & 5.06	& 0.21	 & 41.15	 & 4.65	 & 0.49	& 38.28	 & 4.66	 &   0.05  &  56.87	 & 20.43	  &  0.58 & 49.38	 &  18.29	  &  1.09  &  45.65	& 18.43  \\                              
                                                            
     &      &         &          &  0.68	  & 30.30	  & 3.43	& 0.58	 & 25.85	 & 2.41	 & 0.44	& 20.82	 & 1.48	 &   1.33  &  33.63	 & 11.63	  &  1.15 & 28.59	 &  8.19	  &  0.90  &  22.75	& 4.91   \\                              
                                                            
     &      &         &          &  0.19	  & 46.55	  & 5.04	& 0.17	 & 41.88	 & 4.71	 & 0.48	& 38.78	 & 4.70	 &   0.37  &  57.22	 & 126.31	  &  1.15	 & 50.32	 &  114.65  &  2.62  &  45.63	& 113.44 \\                            
                                                            
     &      &         &          &  0.64	  & 29.89	  & 3.26	& 0.55	 & 25.60	 & 2.32	 & 0.45	& 20.83	 & 1.49	 &   3.16  &  32.88	 & 67.97	  &  2.79	 & 28.23	 &  49.17	  &  2.30  &  22.88	& 31.31  \\ \cmidrule(lr{.25em}){4-4} \cmidrule(lr{.25em}){5-13} \cmidrule(lr{.25em}){14-22}
                                                            
     &      &         &$10^{-3}$ &  0.76	  & 38.72	  & 5.54	& 1.38	 & 34.95	 & 6.87	 & 1.73	& 30.71	 & 7.29	 &   1.48  &  46.67	 & 8.71	    &  2.10 & 40.65	 & 10.57	  &  2.49  & 35.23	  & 11.44  \\                            
                                                            
     &     &  &                  &  1.84	  & 26.58	  & 6.68  & 1.93	 & 23.19	 & 6.30	 & 1.93	& 20.90  & 5.81	 &   2.40  &  28.81  & 9.15	    &  2.47 & 24.71	 & 8.64	    &  2.46  & 22.29	  & 8.13   \\                            
                                                            
     &      &         &          &  0.83	  & 38.48	  & 5.72	& 1.38	 & 33.55	 & 6.48	 & 1.76	& 30.81	 & 7.45	 &   3.05  &  46.15	 & 35.33	  &  4.26	 & 39.08	 & 41.11	  &  4.94  & 35.14	  & 45.13  \\                          
                                                            
     &      &         &          &  1.80	  & 26.35	  & 6.47	& 1.89	 & 23.43	 & 6.20	 & 1.92	& 20.74	 & 5.76	 &   4.68  &  28.27	 & 34.79	  &  4.88	 & 25.32	 & 34.44	  &  4.91  & 21.99	  & 32.12  \\                          
                                                            
     &      &         &          &  0.77	  & 38.63	  & 5.57	& 1.32	 & 33.96	 & 6.36	 & 1.71	& 30.66	 & 7.16	 &   7.69  &  47.46	 & 231.81   & 10.21	 & 39.33	 & 245.49	  & 12.31  &  35.57	& 283.43 \\                            
                                                            
     &      &         &          &  1.84	  & 26.76	  & 6.72	& 1.88	 & 23.52	 & 6.13	 & 1.93	& 20.86	 & 5.82	 &  11.90  &  28.62	 & 224.98   & 12.11	 & 25.08	 & 211.47	  & 12.35  & 22.24	  & 204.18 \\ \cmidrule(lr{.25em}){4-4} \cmidrule(lr{.25em}){5-13} \cmidrule(lr{.25em}){14-22}
                                                            
      &     &  &$5\times 10^{-3}$&  1.90	  & 31.41	  & 25.32	& 2.66	 & 26.17	 & 29.23 & 3.39	& 19.56	 & 41.10 &   5.69  &  35.86	 & 100.86	  &  6.88  & 30.02	 &  115.62	&  7.95  &  22.52	& 151.55 \\                            
                                                            
     &      &         &          &  3.52	  & 18.38	  & 38.51	& 3.97	 & 14.84	 & 42.14 & 4.21	& 12.92	 & 44.56 &   7.56  &  20.03	 & 124.35	  &  8.19  & 16.09	 &  133.49	&  8.49  &  14.03	& 140.03 \\                            
                                                            
     &      &         &          &  2.23	  & 29.07	  & 28.14	& 2.74	 & 24.23	 & 32.54 & 3.19	& 22.56	 & 43.42 &  12.41  &  33.54	 & 441.66	  & 13.86  & 27.47	 &  498.89	& 15.01  &  25.64	& 632.03 \\                            
                                                            
     &      &         &          &  3.46	  & 18.40	  & 38.25	& 3.89	 & 15.32	 & 40.99 & 4.24	& 12.72	 & 44.41 &  14.95  &  19.77	 & 490.83	  & 16.12  & 16.65	 &  522.42	& 17.09  &  13.67	& 558.04 \\                            
                                                            
     &      &         &          &  2.07	  & 29.49	  & 19.21	& 2.64	 & 23.50	 & 25.52 & 3.26	& 21.64	 & 36.30 &  30.16  &  34.71	 & 1959.53  & 34.69  & 27.82	 &  2571.16	& 38.22  &  23.54	& 3357.49\\                            
                                                            
     &      &         &          &  3.45	  & 19.05	  & 37.62	& 3.88	 & 15.45	 & 41.52 & 4.19	& 13.21	 & 44.84 &  37.37  &  20.75	 & 3047.61  & 40.32  & 16.61	 &  3304.94	& 42.35  &  14.20	& 3523.39\\ 
\bottomrule
\end{tabular}                                                                                                                  
\end{table*}

\subsection{Test Sizes in Contaminated Data}\label{sec:simul2}

\begin{figure*}[htb]                                                                                                                                                      
\centering
\subfigure[$\epsilon=0$ \label{figcontamination11}]{\includegraphics[width=.4\linewidth]{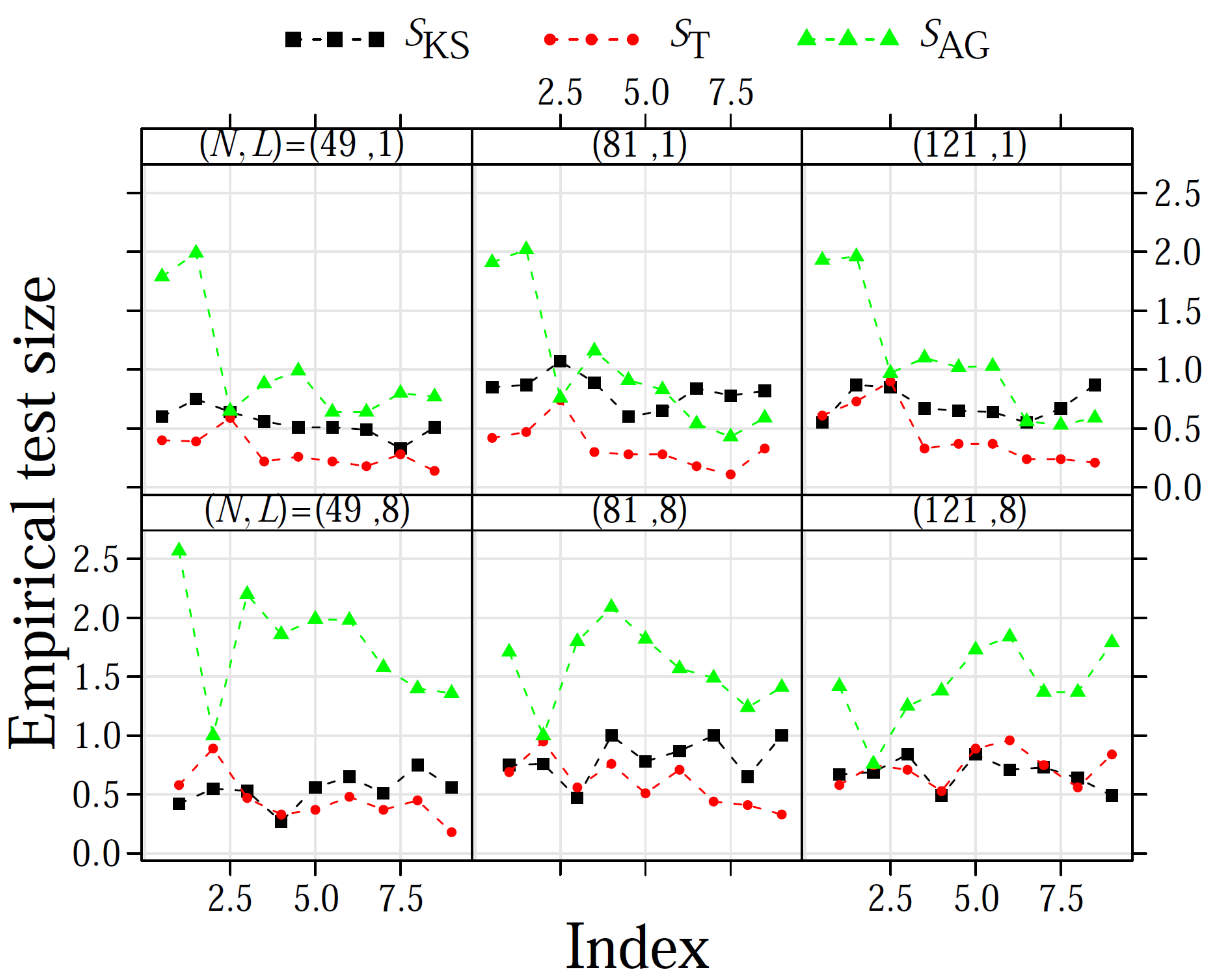}}
\subfigure[$\epsilon=10^{-4}$ \label{figcontamination12}]{\includegraphics[width=.4\linewidth]{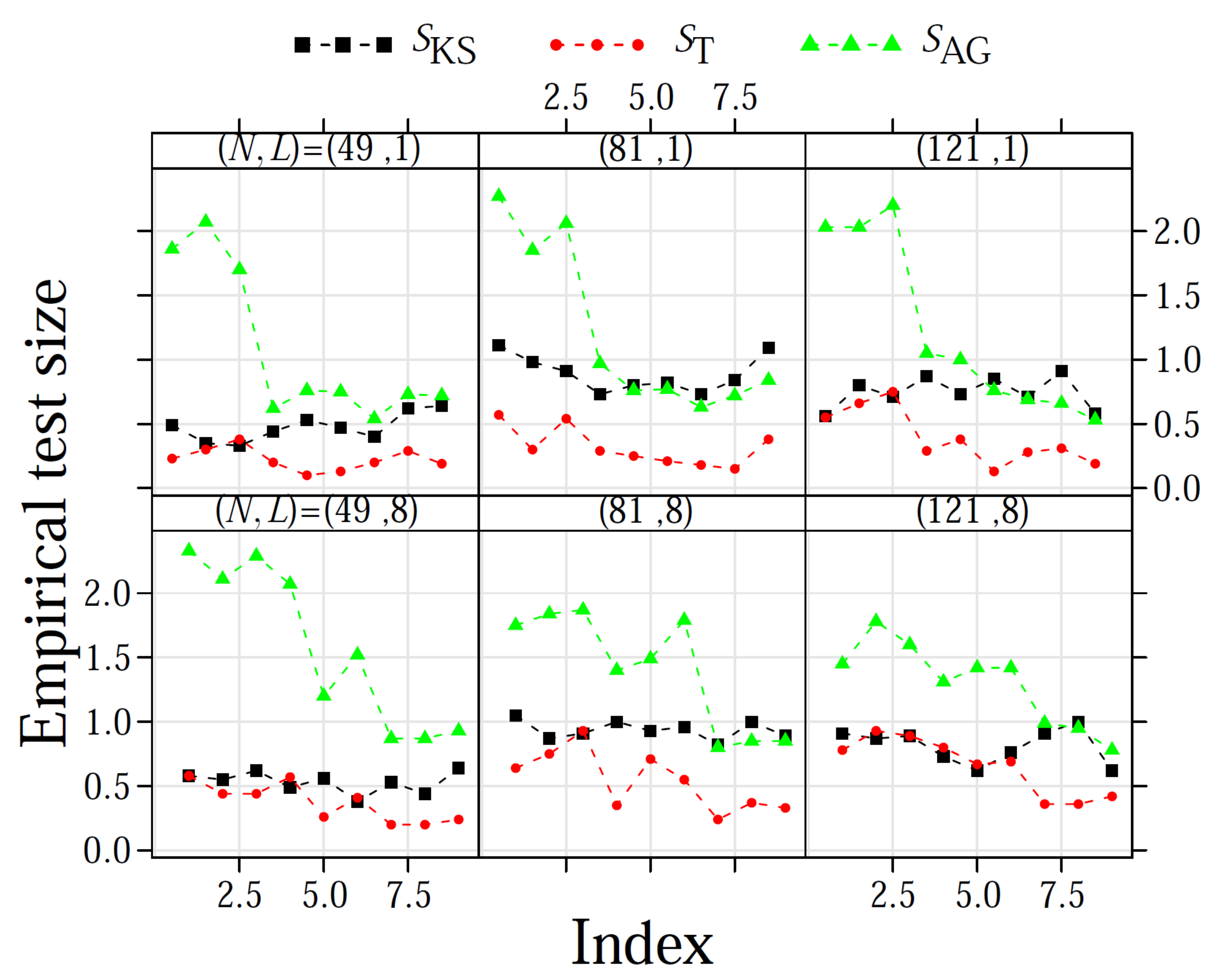}}\\
\subfigure[$\epsilon=5 \times 10^{-3}$ \label{figcontamination14}]{\includegraphics[width=.4\linewidth]{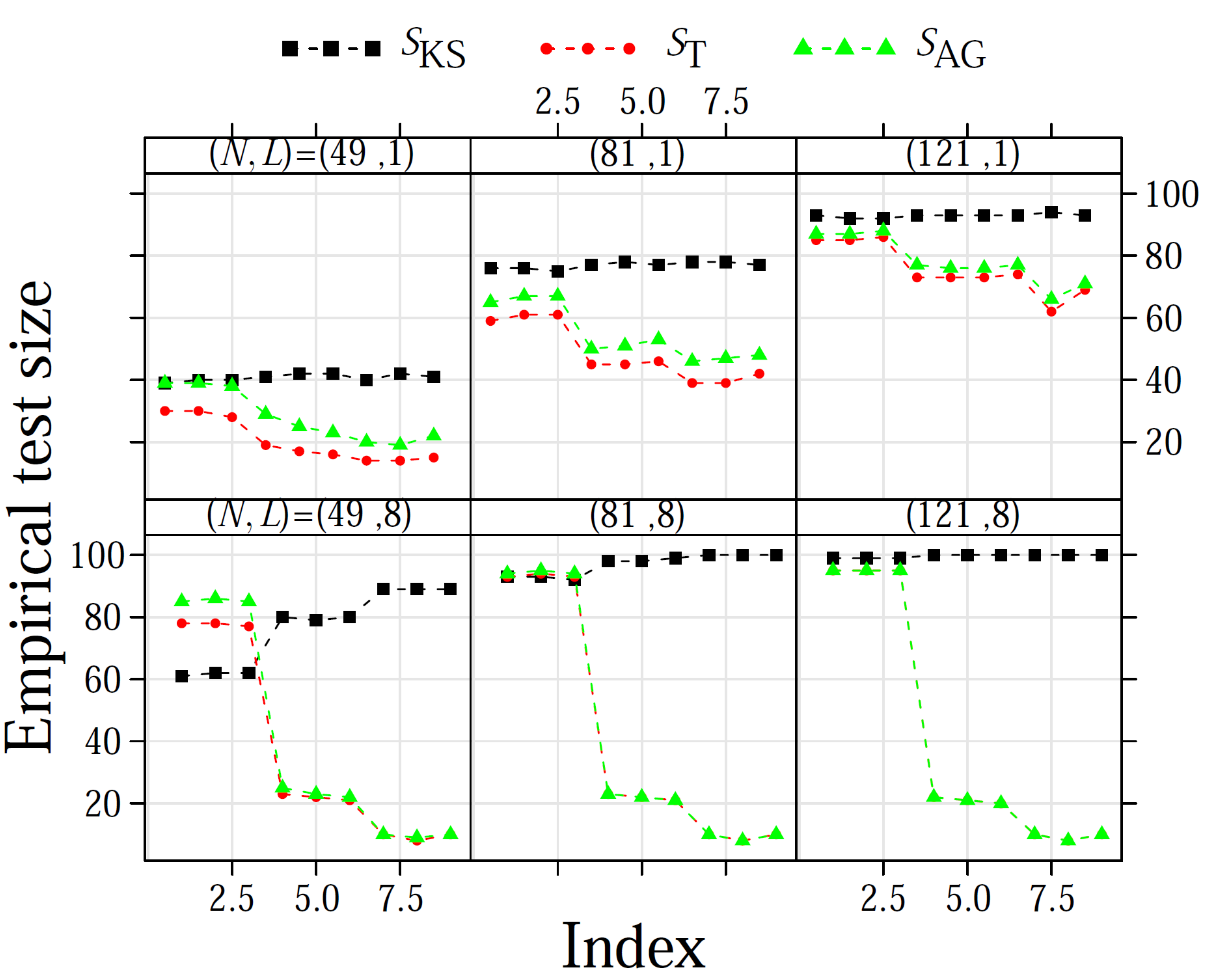}}
\caption{Test sizes at $1\%$ nominal level and $\epsilon$.
For each sample size, all considered scenarios were lexicographically ordered: $[i \colon (i;\alpha,\mu)=\{(1;-1.5,1),\ldots,(4;-3,1),\ldots,(7;-5,1),\ldots\}]$.
Index $i$ was employed as abscissa value.}
\label{figcontamination1}
\end{figure*}

Based on the above results, the forthcoming robustness study is limited to the best statistics, namely $S_\text{T}$ and $S_\text{AG}$, along with $S_\text{KS}$, which is expected to be robust.
For these measures, Fig.~\ref{figcontamination1} shows the behavior of the test size in the presence of fixed levels $\epsilon$ of contamination.

Under contamination, larger windows imply better empirical test sizes for the parametric tests.
Tests based on $S_\text{T}$ are the most robust for homogeneous regions.
The empirical test sizes of $(h,\phi)$-distance based tests for $\alpha=-1.5$ are greater than the ones observed when $\alpha\in\{-3,-5\}$; this behavior is more pronounced as  $\epsilon$ increases.

Unexpectedly, the Kolmogorov-Smirnov test tends to be less robust than parametric tests as contamination increases.
Fig.~\ref{figcontamination1} indicates a low variability of the $S_\text{KS}$ based test size across  different simulation specifications. 
This suggests that image roughness and brightness do not significantly affect these particular estimated test sizes.

\section{Results}\label{sec:real}

\begin{figure}[htb]
\centering
\includegraphics[width=.7\linewidth]{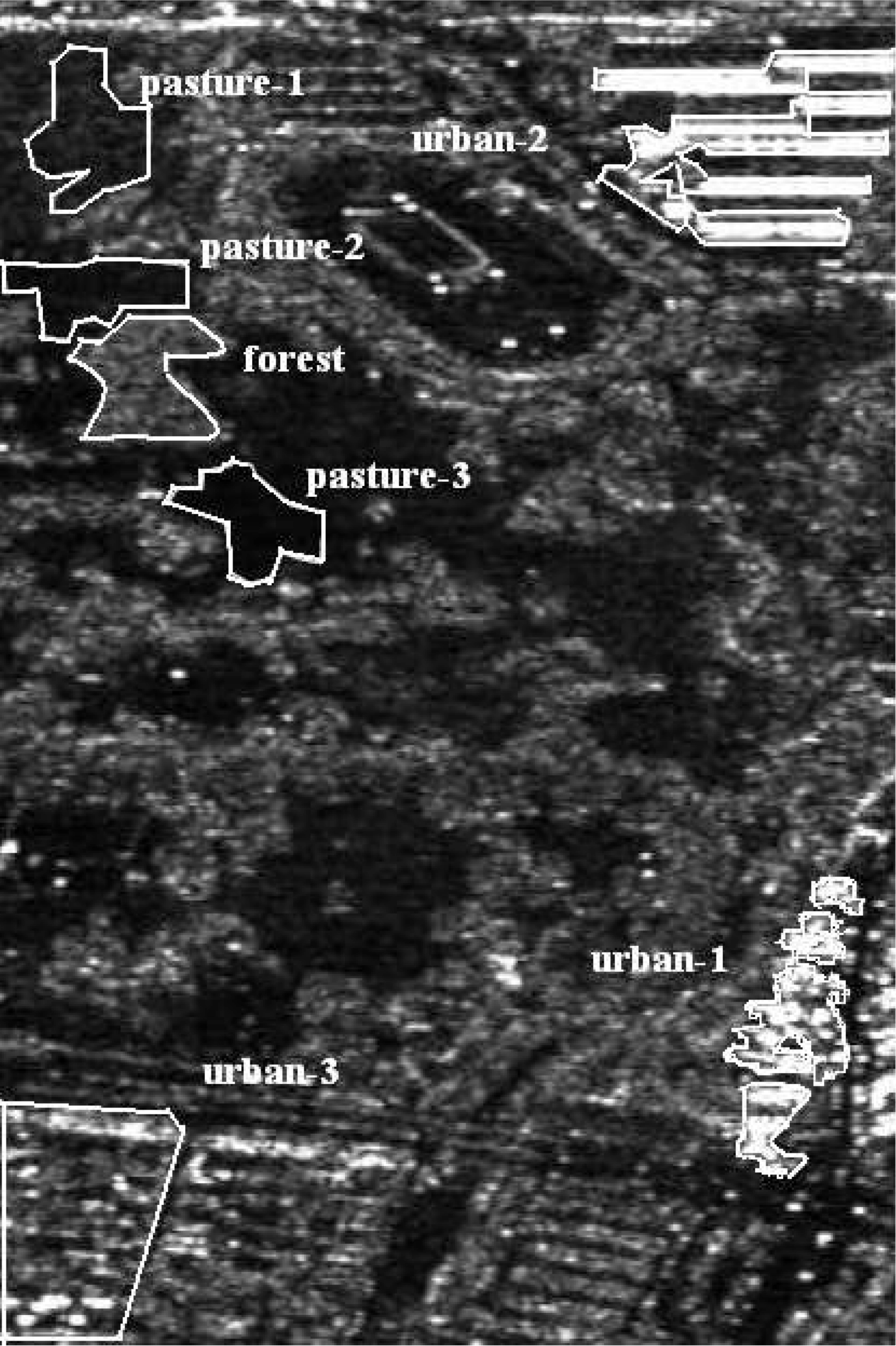}
\caption{E-SAR image and selected regions.}\label{figapplication1}
\end{figure}

In the following we present an application of the proposed tools to real data.
The performance of the considered hypothesis tests are quantified by means of their empirical sizes and powers.
Based on Fig.~\ref{figapplication1}, samples of same and different classes are constrasted.

The image was obtained by the E-SAR sensor over surroundings of Munich, Germany~\cite{ESAR}, and its estimated number of looks is $3.2$.
The area exhibits three distinct types of target roughness: (i)~homogeneous (corresponding to pasture), (ii)~heterogeneous (forest), and (iii)~extremely heterogeneous (urban areas).
Samples were selected and submitted to statistical analysis, after parameter estimation by maximum likelihood methods.

Table~\ref{tabelapplica} shows the $\widehat{\alpha}$, $\widehat{\gamma}$, and $\widehat{\mu}$ estimates (whose interpretability was discussed in Section~\ref{sec:simul}), along with the sample size (\# pixels).
Values of $\widehat{\mu}$ for urban regions are infinite in accordance to the theoretical background as seen in Equation~\eqref{modelmultiplicative4}.
The table also presents the number of disjoint $7 \times 7$ pixels blocks (\#~parts) which were used to estimate the test sizes and powers.

The estimated roughness and brightness parameters satisfied the following inequalities:
$\alpha_{\text{pasture-1}} < \alpha_{\text{pasture-2}} < \alpha_{\text{pasture-3}} < \alpha_{\text{forest}} < \alpha_{\text{urban-3}} < \alpha_{\text{urban-1}} < \alpha_{\text{urban-2}}$ and  $\mu_{\text{pasture-1}} < \mu_{\text{pasture-2}} < \mu_{\text{pasture-3}} < \mu_{\text{forest}} < \mu_{\text{urban-3}}$.
Those inequalities confirm the interpretability of the parameters, and show how hard it is to discern different samples from the same class.

\begin{table}[htb]
\centering
\caption{Parameter estimates}\label{tabelapplica}
\begin{tabular}{c r@{.}l r  r@{.}l  cc}
\toprule
Regions & \multicolumn{2}{c}{$\widehat{\alpha}$} & \multicolumn{1}{c}{$\widehat{\gamma}$} & \multicolumn{2}{c}{$\widehat{\mu}$} & \# pixels & \# parts \\ \midrule

pasture-1 &  $-$15&702 & 39259 &2670&32 & 1235 & 25 \\

pasture-2 &  $-$12&698 & 80320 &6866&13 & 1216  & 24 \\

pasture-3 &  $-$11&304 & 162292 &15750&39 & 1602 & 32 \\

forest  &  $-$9&339 & 661183  &79288&04 & 1606 & 32\\

urban-1   &   $-$0&759 & 148413 &\multicolumn{2}{c}{$\infty$}& 2005 & 40\\

urban-2   &   $-$0&388 & 110183 &\multicolumn{2}{c}{$\infty$}& 3481  & 71 \\

urban-3   &   $-$1&079 & 55583 &703582&30 & 4657 & 95\\
\bottomrule
\end{tabular}
\end{table}

Table~\ref{tabaplic2} presents the rejection rates at significance levels $1\%$ and $10\%$ of samples from the same area, i.e., the test size or Type~I error.
Such nominal levels were considered useful in practical applications (cf. reference~\cite{SilvaCribariFrery:ImprovedLikelihood:Environmetrics}.)
These results show that $S_\text{KS}$, $S_\text{KL}$, and $S_\text{T}$ tests have excellent performance with respect to this criterion, and that the hardest situation is related to urban areas.
Furthermore, the $S_\text{AG}$ test shows optimal size test for $1\%$ nominal level, failing on urban-1 and urban-2 situations at $10\%$, while the $S_\text{B}$ test size is inappropriate in case \mbox{urban-2}.

\begin{table}[htb]
\centering
\scriptsize
\caption{Rejection rates of $(h,\phi)$-divergence tests under $H_1\colon (\alpha_1,\gamma_1)\neq (\alpha_2,\gamma_2)$,
with $\mu_1 =\mu_2$ at $1\%$ and $10 \%$ nominal levels} \label{tabaplic2}
\begin{tabular}{@{ }c r@{ }r@{ }r@{ }r@{ }r r@{ }r@{ }r@{ }r@{ }r@{ }}\toprule
\multicolumn{1}{l}{}& \multicolumn{5}{c}{$1\%$ nominal level} & \multicolumn{5}{c}{$10\%$ nominal level} \\ 
\cmidrule(lr{.25em}){2-6} \cmidrule(lr{.25em}){7-11}
\text{Regions}& $S_\text{KS}$ & $S_\text{KL}$ & $S_\text{T}$ &  $S_\text{B}$&   $S_\text{AG}$ &  $S_\text{KS}$ &
$S_\text{KL}$ & $S_\text{T}$ &  $S_\text{B}$& $S_\text{AG}$ \\ 
\cmidrule(lr{.25em}){1-1} \cmidrule(lr{.25em}){2-6} \cmidrule(lr{.25em}){7-11}
\text{pasture-1} & 0.33 & 0.00 & 0.00 & 0.00  & 0.00 & 5.00 & 0.00  & 0.00  & 0.00  & 0.00  \\

\text{pasture-2} & 0.36 & 0.00 & 0.00 & 0.00  & 0.00 & 6.16 & 0.00  & 0.00  & 0.00  & 0.00  \\

\text{pasture-3} & 0.60 & 0.00 & 0.00 & 0.36  & 0.36 & 6.05 & 3.62  & 3.62  & 3.62  & 3.99  \\
\cmidrule(lr{.25em}){1-1} \cmidrule(lr{.25em}){2-6} \cmidrule(lr{.25em}){7-11}
\text{forest}    & 2.22 & 0.00 & 0.00 & 0.00  & 0.00 & 9.68 & 2.67  & 1.67  & 2.00  & 2.33  \\
\cmidrule(lr{.25em}){1-1} \cmidrule(lr{.25em}){2-6} \cmidrule(lr{.25em}){7-11}
\text{urban-1}   & 0.00 & 0.00 & 0.00 & 0.13  & 0.00 & 2.18 & 4.10  & 2.18  & 34.49 & 90.64 \\

\text{urban-2}   & 0.00 & 0.00 & 0.00 & 99.96 & 0.00 & 0.97 & 2.01  & 1.21  & 99.96 & 100.00   \\

\text{urban-3}   & 0.29 & 4.23 & 1.97 & 3.56  & 5.64 & 5.98 & 20.58 & 16.84 & 20.85 & 31.02 \\ \bottomrule

\end{tabular}
\end{table}

\begin{figure}[htb]
\centering
\includegraphics[width=.8\linewidth]{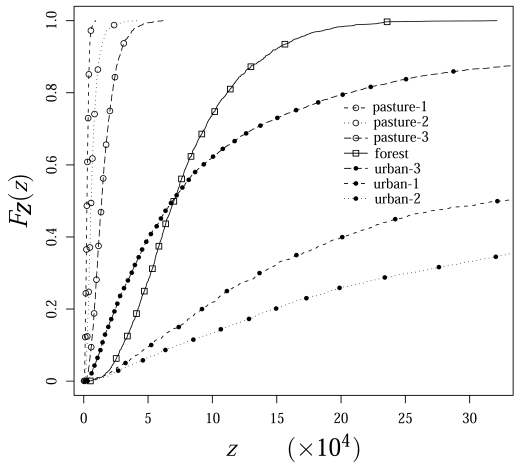}
\caption{Empirical cumulative distribution function for different regions in SAR data for sample values}
\label{cdf}
\end{figure}

Table~\ref{tabaplic4} shows the test powers at $1\%$ and $10\%$ nominal levels, i.e., the inability to reject samples from different types.
The best performances are observed in ($h,\phi$)-divergence tests, mainly when confronting areas whose distributions differ more significantly; among them, $S_\text{T}$ and $S_\text{KL}$ are the weakest ones.
The worst result was observed in the urban-1 vs. urban-2 situation whose samples are, in fact, very close and hard to discern.
The Kolmogorov-Smirnov test produces unacceptable results when contrasting urban-3 vs. forest, and urban-1 vs. urban-2 samples.
In particular, the situation urban-3 vs. forest contrasts two noticeably distinct regions, as shown in Fig.~\ref{figapplication1} and Table~\ref{tabelapplica}.
Fig.~\ref{cdf} presents the empirical distribution functions of the areas considered.
It illustrates the close connection of the $S_\text{KS}$ efficiency to these functions, regardless the texture differences, which makes the $S_\text{KS}$ test an inadequate tool for discrimination in speckled imagery.

\begin{table*}[htb]
\centering
\scriptsize
\caption{Rejection rates of $(h,\phi)$-divergence tests under $H_1\colon (\alpha_1,\gamma_1)\neq (\alpha_2,\gamma_2)$,
 with $\mu_1 =\mu_2$ at $1\%$ and $10\%$ nominal levels}
\label{tabaplic4}
\begin{tabular}{c r@{ }r@{ }r@{ }r@{ }r r@{ }r@{ }r@{ }r@{ }r}\toprule
\multicolumn{1}{l}{}& \multicolumn{5}{c}{$1\%$ nominal level} & \multicolumn{5}{c}{$10\%$ nominal level} \\ 
\cmidrule(lr{.25em}){2-6} \cmidrule(lr{.25em}){7-11}

\text{Regions}& $S_\text{KS}$ & $S_\text{KL}$ & $S_\text{T}$ &  $S_\text{B}$&   $S_\text{AG}$
& $S_\text{KS}$ & $S_\text{KL}$ & $S_\text{T}$ &  $S_\text{B}$&   $S_\text{AG}$ \\ 
\cmidrule(lr{.25em}){1-1} \cmidrule(lr{.25em}){2-6} \cmidrule(lr{.25em}){7-11}

\text{pasture-1$\times$pasture-2}& 100.00   & 100.00   &  100.00   &  100.00   &  100.00      & 100.00   & 100.00   & 100.00   & 100.00   & 100.00    \\

\text{pasture-1$\times$pasture-3}& 100.00   & 100.00   &  100.00   &  100.00   &  100.00      & 100.00   & 100.00   & 100.00   & 100.00   & 100.00    \\

\text{pasture-2$\times$pasture-3}& 100.00   & 94.64 &  87.50 &  90.48 &  95.24    & 100.00   & 98.21 & 97.02 & 98.21 & 98.21  \\
\cmidrule(lr{.25em}){1-1} \cmidrule(lr{.25em}){2-6} \cmidrule(lr{.25em}){7-11}
\text{forest$\times$pasture-1}   & 100.00   & 100.00   &  100.00   &  100.00   &  100.00      & 100.00   & 100.00   & 100.00   & 100.00   & 100.00    \\

\text{forest$\times$pasture-2}   & 100.00   & 100.00   &  100.00   &  100.00   &  100.00      & 100.00   & 100.00   & 100.00   & 100.00   & 100.00    \\

\text{forest$\times$pasture-3}   & 100.00   & 100.00   &  100.00   &  100.00   &  100.00      & 100.00   & 100.00   & 100.00   & 100.00   & 100.00    \\
\cmidrule(lr{.25em}){1-1} \cmidrule(lr{.25em}){2-6} \cmidrule(lr{.25em}){7-11}
\text{urban-1$\times$pasture-1}  & 100.00   & 100.00   &  100.00   &  100.00   &  100.00      & 100.00   & 100.00   & 100.00   & 100.00   & 100.00    \\

\text{urban-1$\times$pasture-2}  & 100.00   & 100.00   &  100.00   &  100.00   &  100.00      & 100.00   & 100.00   & 100.00   & 100.00   & 100.00    \\

\text{urban-1$\times$pasture-3}  & 100.00   & 100.00   &  100.00   &  100.00   &  100.00      & 100.00   & 100.00   & 100.00   & 100.00   & 100.00    \\

\text{urban-2$\times$pasture-1}  & 100.00   & 100.00   &  100.00   &  100.00   &  100.00      & 100.00   & 100.00   & 100.00   & 100.00   & 100.00    \\

\text{urban-2$\times$pasture-2}  & 100.00   & 100.00   &  100.00   &  100.00   &  100.00      & 100.00   & 100.00   & 100.00   & 100.00   & 100.00    \\

\text{urban-2$\times$pasture-3}  & 100.00   & 100.00   &  100.00   &  100.00   &  100.00      & 100.00   & 100.00   & 100.00   & 100.00   & 100.00    \\

\text{urban-3$\times$pasture-1}  & 100.00   & 100.00   &  100.00   &  100.00   &  100.00      & 100.00   & 100.00   & 100.00   & 100.00   & 100.00    \\

\text{urban-3$\times$pasture-2}  & 100.00   & 100.00   &  100.00   &  100.00   &  100.00      & 100.00   & 100.00   & 100.00   & 100.00   & 100.00    \\

\text{urban-3$\times$pasture-3}  & 100.00   & 100.00   &  100.00   &  100.00   &  100.00      & 100.00   & 100.00   & 100.00   & 100.00   & 100.00    \\

\text{urban-1$\times$forest}     & 100.00   & 98.70 &  98.10 &  99.70 &  98.40    & 100.00   & 98.90 & 99.40 & 100.00   & 100.00    \\

\text{urban-2$\times$forest}     & 100.00   & 98.03 &  99.49 &  100.00   &  97.30    & 100.00   & 98.03 & 99.49 & 100.00   & 100.00    \\

\text{urban-3$\times$forest}   & 12.17 & 94.74 &  80.17 &  91.79 &  96.17    & 61.78 & 98.65 & 96.93 & 98.40 & 97.14  \\

\text{urban-1$\times$urban-2}    & 32.04 & 29.68 &  26.37 &  94.19 &  30.60    & 83.87 & 53.20 & 51.73 & 99.47 & 100.00    \\

\text{urban-2$\times$urban-3}    & 98.58 & 91.05 &  89.92 &  94.26 &  91.50    & 100.00   & 96.26 & 96.13 & 97.29 & 98.53  \\

\text{urban-1$\times$urban-3}    & 100.00   & 98.84 &  98.50 &  100.00   &  99.01    & 100.00   & 99.84 & 99.79 & 100.00   & 100.00    \\ \bottomrule

\end{tabular}
\end{table*}

\section{Conclusions}\label{sec:Conc}

This paper presented a comparison among the performance of four parametric tests based on stochastic distances and the Kolmogorov-Smirnov test.
We also presented compact formulas for the Kolmogorov-Smirnov contrast measure, and a study of the robustness of the maximum likelihood estimator under the $\mathcal G^0$ law.
The assessment was made with Monte Carlo experiments varying the parameters of the $\mathcal{G}^0$ distribution, regarded as a universal model for speckled data, and the level of innovative contamination.
In doing so, we provided new results regarding the both the accuracy and the robustness of the measures under examination.

We show numerical evidence that the test based on the triangular distance $S_\text{T}$ has, in general, smaller empirical Ty\-pe~I error than the test based on the Kol\-mo\-go\-rov-Smirnov distance $S_\text{KS}$.
Using a parametric and plausible contamination model, we illustrate that when the chances of observing aberrant data  increase, the test based on $S_\text{KS}$ has its performance diminished with respect to the results obtained from the $(h,\phi)$-distance methods.
This is an important war\-ning due to the widespread use of the Kol\-mo\-go\-rov-Smir\-nov test when dealing with non-normal data.

The $S_\text{AG}$ measure presented the best performance with respect to the test power.
However, for a given number of looks, we observed that the test power performance of the four proposed parametric measures was roughly the same, and that they were more accurate than the test related to the $S_\text{KS}$ measure.

These conclusions suggest the use of the $S_\text{T}$ for SAR data processing due to its resulting accurate test empirical size.
On the other hand, the $S_\text{AG}$ test is a more appropriate tool whenever the focus is on the test power.

We illustrated  the discriminatory capability of these measures using real data.
Tests based on the classical measures here tested, namely the Kullback-Leibler, Kol\-mo\-go\-rov-Smir\-nov, and Bhattacharyya distances, were outperformed by the other distances and should not be employed.

\appendix
\section{Computational information}

Simulations of the ($h,\phi$)-distance based parametric tests were performed in \texttt{Ox} programming environment~\cite{Doornik98}; function \texttt{QAGI} was employed for numerical integration. 
The  Kolmogorov-Smirnov test was coded in \texttt{R} and  function \texttt{ks.test}  was called~\cite{Marsaglia:Tsang:Wang:2003:JSSOBK:v08i18}.
The computation time of the triangular and arithmetic-geometric distances took ty\-pi\-ca\-lly less than one and four millisecond, respectively, when performed in a Pentium processor at~\unit[3.20]{GHz}.
We used the George Marsaglia's multiply-with-carry with 52 bits pseu\-do\-ran\-dom number ge\-ne\-ra\-tor, which has an approximate period of $2^{8222}$.

\bibliographystyle{spmpsci}
\bibliography{Cintraetal}

\end{document}